\documentstyle[12pt]{article}

\newcounter{eq}
\newcounter{sc}

\def\overleftrightarrow#1{\vbox{\ialign{##\crcr
 $\leftrightarrow$\crcr\noalign{\kern-1pt\nointerlineskip}
 $\hfil\displaystyle{#1}\hfil$\crcr}}}

\newlength{\minitwocolumn}
\begin{document}

\begin{flushright}
DPUR/TH/81\\
October, 2024\\
\end{flushright}
\vspace{20pt}

\pagestyle{empty}
\baselineskip15pt

\begin{center}
{\large\bf  BRST Formalism of $f(R)$ Gravity
\vskip 1mm }

\vspace{20mm}

Ichiro Oda\footnote{
           E-mail address:\ ioda@sci.u-ryukyu.ac.jp
                  }

\vspace{10mm}
           Department of Physics, Faculty of Science, University of the 
           Ryukyus,\\
           Nishihara, Okinawa 903-0213, Japan\\

\end{center}


\vspace{10mm}
\begin{abstract}

We perform the manifestly covariant quantization of $f(R)$ gravity in the de Donder gauge condition (or harmonic gauge 
condition) for general coordinate invariance. We explicitly calculate various equal-time commutation relations (ETCRs), 
in particlular, the ETCR between the metric and its time derivative, and show that it has a nonvanishing and nontrivial 
expression, whose situation should be contrasted to the previous result in the higher-derivative or quadratic gravity 
where the ETCR was found to be identically vanishing. We also clarify global symmetries, the physical content of $f(R)$ gravity, 
and clearly show that this theory is manifestly unitary and has a massive scalar and massless graviton as physical modes.

\end{abstract}

\newpage
\pagestyle{plain}
\pagenumbering{arabic}


\section{Introduction}

It is valuable to seek for various modifications of general relativity by Einstein since quantum theory of general relativity
is known to be nonrenormalizable in the conventional perturbative approach. Recently, $f(R)$ gravity has attracted much 
attention by several reasons, especially from the viewpoint of cosmology \cite{Hwang, Cognola, Capozziello, Sotiriou, Felice}. 
It is also of interest that $f(R)$ gravity belongs to a class of higher-derivative gravity but it is free of ghosts and 
consequently unitary though it is not renormalizable. 
 
Thus far, there have been a lot of studies of the $f(R)$ gravity from both classical and quantum sides, but as long as we know,
there is no study of performing its manifestly covariant quantization in the de Donder gauge condition (or harmonic gauge 
condition) for general coordinate invariance, investigating its global symmetries, and verifying the physical states on the basis of 
the BRST formalism \cite{Kugo-Ojima}. Performing such a manifestly covariant quantization in an explicit way is one of motivations 
in this article.

Another motivation behind this article is to understand the BRST formalism of the higher-derivative gravity, which is 
recently called $\it{quadratic \, gravity}$ as well. The quadratic gravity is known to be renormalizable but is not unitary
owing to the presence of massive ghost \cite{Stelle1} and its Lagrangian takes a symbolic form, 
${\cal{L}}_{QG} = \sqrt{-g} \left( \frac{1}{2 \kappa^2} R + \alpha R^2 + \beta C_{\mu\nu\rho\sigma}^2 \right)$.
In the manifestly covariant quantization procedure of the quadratic gravity, it is mentioned in Refs. \cite{Kimura1, Kimura2, Kimura3} 
that the equal-time commutation relation (ETCR) between the metric and its time derivative identically vanishes, 
$[ g_{\mu\nu}, \dot g_{\rho\sigma}^\prime ] = 0$,\footnote{Using this ETCR and the de Donder gauge condition, 
we can also show $[ \dot g_{\mu\nu}, \dot g_{\rho\sigma}^\prime ] = 0$.}  
whereas it is shown in Refs. \cite{Nakanishi, N-O-text} that in general relativity this ETCR is nonvanishing. 
Of course, this vanishing ETCR does not imply that the gravitational modes become trivial in the quadratic gravity 
since dynamical degrees of freedom associated with the graviton and the other modes are involved in the auxiliary fields 
such as the Nakanishi-Lautrup field and the Stuckelberg field. Nevertheless, this vanishing ETCR not only enables us 
to easily carry out complicated calculations of the ETCRs relevant to the metric field, but also might exhibit an interesting 
feature of the quadratic gravity such as a sort of triviality of the gravitational sector at high energies 
where the higher-derivative terms are more dominant than the Einstein-Hilbert term. Furthermore, the ETCR, 
$[ g_{\mu\nu}, \dot g_{\rho\sigma}^\prime ] = 0$ means that the canonical formalism of the quadratic gravity is not smoothly 
connected with that of general relativity in the suitable limit $\alpha, \beta \rightarrow 0$.

In this article, we would like to verify whether the vanishing ETCR is also valid or not in a simpler theory where the conformal 
tensor-squared term, $C_{\mu\nu\rho\sigma}^2$ is dropped and an $R^2$ term is replaced with the $f(R)$ term 
in the quadratic gravity, that is, the Lagrangian is of form, ${\cal{L}}_0 = \sqrt{-g} \left( \frac{1}{2 \kappa^2} R + f(R) \right)$ 
since the BRST quantization of the $C_{\mu\nu\rho\sigma}^2$ term requires us to treat a rather complicated Stuckelberg symmetry 
\cite{Kimura1, Kimura2, Kimura3}.  We will see that the ETCR, $[ g_{\mu\nu}, \dot g_{\rho\sigma}^\prime ]$ in $f(R)$ gravity
is nonvanishing and has a more complicated expression than that of general relativity. This observation might imply that 
the $C_{\mu\nu\rho\sigma}^2$ term might play a role for making the ETCR be vanishing in the quadratic gravity.  
In fact, in conformal gravity described by the Lagrangian,  ${\cal{L}}_{C^2} = \sqrt{-g} \, \beta C_{\mu\nu\rho\sigma}^2$, 
it has been already known that the ETCR, $[ g_{\mu\nu}, \dot g_{\rho\sigma}^\prime ]$ has a rather simpler expression 
compared to that of general relativity \cite{Oda-Ohta}.
Finally we wish to refer to related articles \cite{Oda-Q, Oda-W, Oda-Saake, Oda-Conf}.

We close this section with an overview of this article. In Section 2, we review a relation between $f(R)$ gravity and 
scalar-tensor gravity. In Section 3, we construct a BRST-invariant action, derive field equations and argue its global symmetries.
In Section 4, starting with the canonical commutation relations we calculate various ETCRs where a special attention 
is paid for the ETCR between the metric and its time derivative. In Section 5, we clarify the physical content by imposing the Kugo-Ojima 
condition \cite{Kugo-Ojima} and show that the physical S-matrix is unitary in the theory at hand . The final section is devoted to 
the conclusion. 
Two appendices are put for technical details. In Appendix A, two different derivations of the field equation for $b_\mu$ field
are given, and in Appendix B we present two kinds of different derivations of $[ b_\mu, \dot b_\nu^\prime ]$.

\section{Relation between $f(R)$ gravity and scalar-tensor gravity}

We wish to consider a model of $f(R)$ gravity defined by the Lagrangian density
\begin{eqnarray}
{\cal{L}}_0 = \sqrt{-g} \left[ \frac{1}{2 \kappa^2} R + f(R) \right],
\label{L-0}  
\end{eqnarray}
where $\kappa$ is defined as $\kappa^2 = 8 \pi G = \frac{1}{M_{Pl}^2}$ through the Newton constant $G$ and the
reduced Planck mass $M_{Pl}$.\footnote{We follow the notation and conventions
of MTW textbook \cite{MTW}. Greek little letters $\mu, \nu, \cdots$ and Latin ones
$i, j, \cdots$ are used for space-time and spatial indices, respectively; for instance, $\mu= 0, 1, 2, 3$ and
$i = 1, 2, 3$. For simplicity, henceforth we set $\kappa = 1$.}  Since general relativity nicely describes many of gravitational
and cosmological phenomena at low energies, we have extracted the Einstein-Hilbert term from the $f(R)$ term. This definition 
is slightly different from that of the conventional $f(R)$ gravity where $f(R)$ term includes the Einstein-Hilbert
term.

Since the Lagrangian (\ref{L-0}) possesses higher-derivative terms in the $f(R)$ term, 
it is necessary to transform it into the first-order formalism in the canonical quantization procedure.
For this aim, let us first impose a constraint $\chi = R$ by introducing a Lagrange multiplier field $\phi$:
\begin{eqnarray}
{\cal{L}}_{\phi-\chi} = \sqrt{-g} \left[ \frac{1}{2} R + \phi ( R - \chi ) + f(\chi) \right].
\label{L-1}  
\end{eqnarray}
The two Lagrangians (\ref{L-0}) and (\ref{L-1}) are equivalent at quantum level since in (\ref{L-1})
the path integral over $\phi$ produces the delta functional $\delta ( \chi - R)$ and then integrating over $\chi$  
exactly leads to (\ref{L-0}). 

Next, let us take the variation with respect to $\chi$ in (\ref{L-1}), from which we have the field equation
\begin{eqnarray}
\phi = f^\prime(\chi),
\label{phi-f}  
\end{eqnarray}
where $f^\prime(\chi) \equiv \frac{d f}{d \chi}$. Inserting (\ref{phi-f}) to the Lagrangian (\ref{L-1}), 
we can obtain:
\begin{eqnarray}
{\cal{L}}_{\chi} = \sqrt{-g} \left[ \left( \frac{1}{2} + f^\prime(\chi) \right) R - f^\prime(\chi) \chi + f(\chi) \right].
\label{L-2}  
\end{eqnarray}
It is obvious that the Lagrangian (\ref{L-2}) is equivalent to (\ref{L-0}) at least classically since we have
made use of the field equation for $\chi$. Note that $\chi$ in (\ref{L-2}) is implicitly defined in terms of $\phi$ through 
Eq. (\ref{phi-f}). When the one-variable function $\phi = f^\prime(\chi)$ is not only continuous and 
differentiable at $\chi = R$ but also $f^{\prime\prime}(R) \neq 0$, the inverse function theorem says that
$\phi = f^\prime(\chi)$ can be uniquely solved in the neighborhood of $\chi = R$ and be expressed by\footnote{Provided that
we consider $H(\chi, \phi) \equiv \phi - f^\prime(\chi)$, the inverse function theorem is a specific example of one-parameter
implicit function theorem.}
\begin{eqnarray}
\chi = F(\phi),
\label{F-phi}  
\end{eqnarray}
where $F$ is the inverse function of $f^\prime(\chi)$, that is,
\begin{eqnarray}
F(f^\prime(\chi) ) = \chi, \qquad f^\prime(F(\phi)) = \phi.
\label{F-inverse}  
\end{eqnarray}
Then, substituting Eqs. (\ref{phi-f}) and (\ref{F-phi}) into (\ref{L-2}), we arrive at our classical Lagrangian
which consists of the Einstein-Hilbert term, the non-minimal coupling term and a potential
\begin{eqnarray}
{\cal{L}}_c = \sqrt{-g} \left[ \frac{1}{2} R + \phi R + \Phi(\phi) \right],
\label{Our-L}  
\end{eqnarray}
where we have defined $\Phi(\phi) \equiv - \phi F(\phi) + f(F(\phi))$.  In this article we shall apply a BRST quantization 
method to this Lagrangian (\ref{Our-L}). Compared to the conventional scalar-tensor gravity \cite{Fujii},
the Lagrangian (\ref{Our-L}) contains the Einstein-Hilbert term but does not include the kinetic term for the scalar field, 
$- \sqrt{-g} \, \frac{1}{\phi} \, g^{\mu\nu} \partial_\mu \phi \partial_\nu \phi$ (recall that $\phi$ has a mass dimension $2$)
though it is implicitly provided by the non-minimal coupling term, $\sqrt{-g} \phi R$, as will be seen later. 

To close this section, we would like to comment on two remarks. First, as a classically equivalent Lagrangian to the $f(R)$ gravity,
the Lagrangian (\ref{L-2}) is sometimes utilized. Actually, taking the variation of $\chi$ yields:
\begin{eqnarray}
f^{\prime\prime} (\chi) ( R - \chi ) = 0.
\label{f^2}  
\end{eqnarray}
As long as $f^{\prime\prime} (\chi) \neq 0$, this equation gives us $\chi = R$, and then substituting this relation into
(\ref{L-2}), we recover (\ref{L-0}). Note that the condition $f^{\prime\prime} (\chi) \neq 0$ corresponds to the condition
$f^{\prime\prime}(R) \neq 0$, which is needed for the inverse function theorem. Incidentally, there are some references
which insist that the equivalence between (\ref{L-0}) and (\ref{L-2}) be valid even at the one-loop level on shell
in arbitrary space-time dimensions \cite{Ruf, Ohta}.

As the second remark, when $f(R)$ is quadratic in $R$, that is, $f(R) = \alpha R^2$ with a constant $\alpha$,  
it is possible to rewrite (\ref{L-0}) by introducing a scalar field $\phi$ with mass squared dimension into 
a quantum mechanically equivalent form:
\begin{eqnarray}
{\cal{L}}_{R^2} = \sqrt{-g} \left( \frac{1}{2} R + \phi R - \frac{1}{4 \alpha} \phi^2 \right).
\label{R^2-equiv}  
\end{eqnarray}
Thus, for the quadratic gravity without the conformal tensor squared, the quantum equivalence between the Lagrangian
(\ref{L-0}) and (\ref{Our-L}) (or (\ref{R^2-equiv})) is guaranteed. The Lagrangian (\ref{R^2-equiv}) will be 
considered in Section 5 in analyzing the physical content of the $f(R)$ gravity.

\section{BRST transformation and field equations}

Let us fix the general coordinate symmetry by the de Donder gauge condition
\begin{eqnarray}
\partial_\mu\tilde g^{\mu\nu} = 0,
\label{Donder}  
\end{eqnarray}
where we have defined $\tilde g^{\mu\nu} = \sqrt{-g} g^{\mu\nu}$. Then, the BRST transformation is of
form:
\begin{eqnarray}
\delta_B \bar c_\rho&=& i B_\rho, \quad \delta_B c^\rho= - c^\lambda\partial_\lambda c^\rho,
\quad \delta_B \phi = - c^\lambda \partial_\lambda \phi,
\nonumber\\
\delta_B g_{\mu\nu} &=& - ( \nabla_\mu c_\nu+ \nabla_\nu c_\mu)
\nonumber\\
&=& - ( c^\alpha\partial_\alpha g_{\mu\nu} + \partial_\mu c^\alpha g_{\alpha\nu} 
+ \partial_\nu c^\alpha g_{\mu\alpha} ),
\nonumber\\
\delta_B \tilde g^{\mu\nu} &=& \sqrt{-g} ( \nabla^\mu c^\nu+ \nabla^\nu c^\mu 
- g^{\mu\nu} \nabla_\rho c^\rho).
\label{BRST}  
\end{eqnarray}

Using this BRST transformation, the Lagrangian for the gauge-fixing condition and 
Faddeev-Popov (FP) ghosts can be constructed in a standard manner:
\begin{eqnarray}
{\cal{L}}_{GF + FP} &=& \delta_B ( i \tilde g^{\mu\nu} \partial_\mu\bar c_\nu)
\nonumber\\
&=& - \tilde g^{\mu\nu} \partial_\mu B_\nu- i \partial_\mu\bar c_\nu\Bigl[ \tilde g^{\mu\rho} \partial_\rho c^\nu 
+ \tilde g^{\nu\rho} \partial_\rho c^\mu - \partial_\rho( \tilde g^{\mu\nu} c^\rho) \Bigr]. 
\label{GF+FP}  
\end{eqnarray}
To simplify this expression, let us introduce a new auxiliary field $b_\rho$ defined as
\begin{eqnarray}
b_\rho= B_\rho- i c^\lambda\partial_\lambda\bar c_\rho,
\label{b-field}  
\end{eqnarray}
and its BRST transformation reads:
\begin{eqnarray}
\delta_B b_\rho= - c^\lambda\partial_\lambda b_\rho.
\label{b-BRST}  
\end{eqnarray}
Then, the Lagrangian (\ref{GF+FP}) can be cast to the form:
\begin{eqnarray}
{\cal{L}}_{GF + FP} = - \tilde g^{\mu\nu} \partial_\mu b_\nu- i \tilde g^{\mu\nu} \partial_\mu\bar c_\rho 
\partial_\nu c^\rho+ i \partial_\rho( \tilde g^{\mu\nu} \partial_\mu\bar c_\nu\cdot c^\rho). 
\label{GF+FP2}  
\end{eqnarray}

As a result, up to a total derivative, the gauge-fixed and BRST-invariant quantum Lagrangian is given 
by\footnote{This Lagrangian is also invariant under $GL(4)$ transformation.}
\begin{eqnarray}
{\cal{L}} = \sqrt{-g} \left[ \frac{1}{2} R + \phi R + \Phi(\phi) \right] - \tilde g^{\mu\nu} \partial_\mu b_\nu 
- i \tilde g^{\mu\nu} \partial_\mu\bar c_\rho\partial_\nu c^\rho.
\label{q-Lag}  
\end{eqnarray}
From this Lagrangian, we can obtain field equations by taking the variation with respect to $g_{\mu\nu}$,
$\phi$, $b_\nu$, $\bar c_\rho$ and $c^\rho$, in order:
\begin{eqnarray}
&{}& \left( \frac{1}{2} + \phi \right) G_{\mu\nu} - ( \nabla_\mu\nabla_\nu- g_{\mu\nu} \Box ) \phi 
- \frac{1}{2} g_{\mu\nu} \Phi(\phi) - \frac{1}{2} ( E_{\mu\nu} - \frac{1}{2} g_{\mu\nu} E ) = 0,
\nonumber\\
&{}& R + \Phi^\prime(\phi) = 0,  \qquad \partial_\mu\tilde g^{\mu\nu} = 0,
\nonumber\\
&{}& g^{\mu\nu} \partial_\mu\partial_\nu c^\rho = 0,  \qquad
g^{\mu\nu} \partial_\mu\partial_\nu\bar c_\rho = 0,
\label{Field-eq}  
\end{eqnarray}
where $G_{\mu\nu} \equiv R_{\mu\nu} - \frac{1}{2} g_{\mu\nu} R$ is the Einstein tensor and we have defined 
\begin{eqnarray}
E_{\mu\nu}  &=& \partial_\mu b_\nu+ i \partial_\mu\bar c_\rho\partial_\nu c^\rho+ ( \mu\leftrightarrow \nu),
\nonumber\\
E &=& g^{\mu\nu} E_{\mu\nu}.
\label{E}  
\end{eqnarray}
From the derivations in Appendix A, it turns out that the field equation for $b_\rho$ field takes the form: 
\begin{eqnarray}
g^{\mu\nu} \partial_\mu\partial_\nu b_\rho = 0.
\label{b-field-eq}  
\end{eqnarray}

Hence, together with space-time coordinates $x^\mu$\footnote{It is necessary to incorporate 
the space-time coordinates $x^\mu$ into $X^M$ in order to obtain the global symmetry.}, 
the $b_\rho$ field, the ghost field $c^\rho$ and the antighost field $\bar c_\rho$ all satisfy 
the d'Alembert equation:
\begin{eqnarray}
g^{\mu\nu} \partial_\mu\partial_\nu X^M = 0,
\label{Alembert-eq}  
\end{eqnarray}
where we have denoted $X^M = \{ x^\mu, b_\mu, c^\mu, \bar c_\mu \}$. With the help of the de Donder gauge
condition (\ref{Donder}), this equation provides us with two kinds of conserved currents
\begin{eqnarray}
{\cal{P}}^{\mu M} &=& g^{\mu\nu} \partial_\nu X^M = g^{\mu\nu} ( 1 \overleftrightarrow{\partial}_\nu X^M ),
\nonumber\\
{\cal{M}}^{\mu M N} &=& g^{\mu\nu} ( X^M \overleftrightarrow{\partial}_\nu Y^N ),
\label{IOSp-current}  
\end{eqnarray}
and the corresponding charges
\begin{eqnarray}
{\cal{P}}^{M} = \int d^3 x \, {\cal{P}}^{0 M}, \qquad 
{\cal{M}}^{M N} = \int d^3 x \, {\cal{M}}^{0 M N},
\label{IOSp-charge}  
\end{eqnarray}
where $X^M \overleftrightarrow{\partial}_\mu Y^N \equiv X^M \partial_\mu Y^N -
(\partial_\mu X^M) Y^N$. The charges ${\cal{P}}^{M}$ and ${\cal{M}}^{M N}$ have $16$ and $128$ generators, respectively, 
and they constitute a $16$-dimensional global Poincare-like symmetry $IOSp (8, 8)$ as in general relativity \cite{Nakanishi, N-O-text}.
In other words, the quantum $f(R)$ gravity under consideration possesses a very huge $IOSp (8, 8)$ symmetry.
Of course, the existence of such a global symmetry is not inconsistent with the black hole no-hair theorem \cite{MTW}
since the present global symmetry is of quantum origin. Actually, the whole generator of the $IOSp (8, 8)$ symmetry
is in essence constructed out of quantum fields such as $b_\mu$ field and FP ghosts.

\section{Equal-time commutation relations}

In this section, after introducing canonical commutation relations (CCRs), we wish to evaluate various equal-time 
commutation relations (ETCRs) among fundametal variables in $f(R)$ gravity. In particular, we are interested in
the ETCR between the metric field and its time derivative. In previous works of the quadratic gravity 
\cite{Kimura1, Kimura2, Kimura3}, it was mentioned that in much contrast to the case of Einstein's general 
relativity \cite{Nakanishi, N-O-text}, its ETCR vanishes identically. The $f(R)$ gravity involves higher-derivative terms 
as in the quadratic gravity so it is of interest whether this feature is taken over by the $f(R)$ gravity
or not. We will show clearly that the ETCR between the metric field and its time derivative is not only nonvanishing 
but also has a more complicated structure than that of general relativity.

To simplify various expressions, we follow the abbreviations adopted in the textbook of Nakanishi
and Ojima\footnote{Our notation is $x^\mu = (x^0, x^i) = (t, x^i)$ and $p^\mu = (p^0, p^i) = (E, p^i)$.}:
\begin{eqnarray}
[ A, B^\prime ] &=& [ A(x), B(x^\prime) ] |_{x^0 = x^{\prime 0}},
\qquad \delta^3 = \delta(\vec{x} - \vec{x}^\prime), 
\nonumber\\
\tilde f &=& \frac{1}{\tilde g^{00}} = \frac{1}{\sqrt{-g} g^{00}},
\label{Def}  
\end{eqnarray}
where $\tilde g^{00}$ is assumed to be invertible. 

To remove second-order derivatives of the metric involved in $R$, we perform the integration by parts once and
rewrite the Lagrangian (\ref{q-Lag}) as
\begin{eqnarray}
{\cal{L}} &=& - \left( \frac{1}{2} + \phi \right) \tilde g^{\mu\nu} ( \Gamma^\alpha_{\sigma\alpha} \Gamma^\sigma_{\mu\nu} 
- \Gamma^\alpha_{\sigma\nu} \Gamma^\sigma_{\mu\alpha} ) - \partial_\mu\phi( \tilde g^{\alpha\beta} 
\Gamma^\mu_{\alpha\beta}  - \tilde g^{\mu\nu} \Gamma^\alpha_{\nu\alpha} ) 
\nonumber\\
&+& \sqrt{-g} \Phi(\phi) + \partial_\mu\tilde g^{\mu\nu} b_\nu 
- i \tilde g^{\mu\nu} \partial_\mu\bar c_\rho\partial_\nu c^\rho+ \partial_\mu{\cal{V}}^\mu,
\label{conv-Lag}  
\end{eqnarray}
where a surface term ${\cal{V}}^\mu$ is defined as
\begin{eqnarray}
{\cal{V}}^\mu=  \left( \frac{1}{2} + \phi \right) ( \tilde g^{\alpha\beta} \Gamma^\mu_{\alpha\beta} - \tilde g^{\mu\nu} 
\Gamma^\alpha_{\nu\alpha} ) - \tilde g^{\mu\nu} b_\nu.
\label{surface}  
\end{eqnarray}

For later convenience, let us take account of the de Donder gauge condition (\ref{Donder}), from which we have
identities:
\begin{eqnarray}
g^{\mu\nu} \Gamma^\lambda_{\mu\nu} = 0, \qquad  g^{\lambda\mu} \partial_\lambda g_{\mu\nu} 
= \Gamma^\lambda_{\lambda\nu}.
\label{Donder-iden}  
\end{eqnarray}
These identities and the de Donder gauge condition will be often utilized to arrive at final expressions of
various results.
Moreover, since the equation $g^{\mu\nu} \Gamma^\lambda_{\mu\nu} = 0$ reads 
\begin{equation}
( 2 g^{\lambda\mu} g^{\nu\rho} - g^{\mu\nu} g^{\lambda\rho} )\partial_\rho g_{\mu\nu}=0,
\label{Donder-iden2}
\end{equation}
it is possible to express the time derivative of the metric field 
in terms of its spacial one as
\begin{eqnarray}
{\cal{D}}^{\lambda\mu\nu} \dot g_{\mu\nu} = ( 2 g^{\lambda\mu} g^{\nu k} - g^{\mu\nu} g^{\lambda k} )
\partial_k g_{\mu\nu}, 
\label{D-eq}  
\end{eqnarray}
where the operator ${\cal{D}}^{\lambda\mu\nu}$ is defined by
\begin{eqnarray}
{\cal{D}}^{\lambda\mu\nu} = g^{0 \lambda} g^{\mu\nu} - 2 g^{\lambda\mu} g^{0 \nu}.
\label{D-op}  
\end{eqnarray}

Now let us set up the canonical (anti)commutation relations (CCRs): 
\begin{eqnarray}
[ \pi_g^{\rho\lambda}, g_{\mu\nu}^\prime ] &=& - i \frac{1}{2} ( \delta_\mu^\rho\delta_\nu^\lambda 
+ \delta_\mu^\lambda\delta_\nu^\rho) \delta^3,  \qquad [ \pi_\phi, \phi^\prime ] = - i \delta^3,
\nonumber\\
\{ \pi_{c \lambda}, c^{\sigma\prime} \} &=& \{ \pi_{\bar c}^\sigma,  \bar c_\lambda^\prime \}
= i \delta_\lambda^\sigma\delta^3.
\label{CCR}  
\end{eqnarray}
Here the canonical variables are $g_{\mu\nu}, \phi, c^\rho, \bar c_\rho$ and the corresponding canonical
conjugate momenta are $\pi_g^{\mu\nu}, \pi_\phi, \pi_{c \rho}, \pi_{\bar c}^\rho$, respectively and 
the $b_\mu$ field is regarded as not a canonical variable but a conjugate momentum of $\tilde g^{0 \mu}$. 

Based on the Lagrangian (\ref{conv-Lag}), the concrete expressions for canonical conjugate momenta read:
\begin{eqnarray}
\pi_g^{\mu\nu} &=& \frac{\partial{\cal{L}}}{\partial\dot g_{\mu\nu}} 
\nonumber\\
&=& - \frac{1}{2} \sqrt{-g} \, \left( \frac{1}{2} + \phi \right) \Bigl[ - g^{0 \lambda} g^{\mu\nu} g^{\sigma\tau} 
- g^{0 \tau} g^{\mu\lambda} g^{\nu\sigma} - g^{0 \sigma} g^{\mu\tau} g^{\nu\lambda} 
\nonumber\\
&+& g^{0 \lambda} g^{\mu\tau} g^{\nu\sigma} 
+ g^{0 \tau} g^{\mu\nu} g^{\lambda\sigma}
+ \frac{1}{2} ( g^{0 \mu} g^{\nu\lambda} + g^{0 \nu} g^{\mu\lambda} ) g^{\sigma\tau} \Biggr] \partial_\lambda g_{\sigma\tau}
\nonumber\\
&-& \sqrt{-g} \Biggl[ \frac{1}{2} ( g^{0 \mu} g^{\rho\nu} + g^{0 \nu} g^{\rho\mu} ) - g^{\mu\nu} g^{\rho0} \Bigr] 
\partial_\rho\phi 
\nonumber\\
&-& \frac{1}{2} \sqrt{-g} ( g^{0 \mu} g^{\nu\rho} + g^{0 \nu} g^{\mu\rho} - g^{0 \rho} g^{\mu\nu} )  b_\rho,
\nonumber\\
\pi_\phi&=& \frac{\partial{\cal{L}}}{\partial\dot \phi} = - ( \tilde g^{\alpha\beta} \Gamma^0_{\alpha\beta}
- \tilde g^{0 \nu} \Gamma^\alpha_{\nu\alpha} ),
\nonumber\\
\pi_{c \sigma} &=& \frac{\partial{\cal{L}}}{\partial\dot c^\sigma} = - i \tilde g^{0\mu} \partial_\mu\bar c_\sigma,
\nonumber\\
\pi_{\bar c}^\sigma&=& \frac{\partial{\cal{L}}}{\partial\dot {\bar c}_\sigma} = i \tilde g^{0\mu} \partial_\mu c^\sigma,
\label{CCM}  
\end{eqnarray}
where differentiation of ghosts is taken from the right, and the other (anti)commutation relations vanish. 
Here the dot denotes time derivative, e.g., $\dot{g}_{\mu\nu} = \partial_0 g_{\mu\nu} 
= \frac{\partial g_{\mu\nu}}{\partial x^0} = \frac{\partial g_{\mu\nu}}{\partial t}$. 

From now on, we would like to evaluate various nontrivial equal-time commutation relations (ETCRs).
In particular, we will exhibit that the equal-time commutation relation between the metric field and its time derivative 
takes a nonvanishing and complicated expression.

For this purpose, let us start with the CCR:
\begin{eqnarray}
[ \pi_g^{\alpha0}, g_{\mu\nu}^\prime ] = - i \frac{1}{2} ( \delta_\mu^\alpha\delta_\nu^0 
+ \delta_\mu^0 \delta_\nu^\alpha) \delta^3.
\label{pi(a0)-g}  
\end{eqnarray}
The canonical conjugate momentum $\pi_g^{\alpha0}$ has a structure
\begin{eqnarray}
\pi_g^{\alpha0} = A^\alpha+ B^{\alpha\beta} \partial_\beta\phi+ C^{\alpha\beta} b_\beta,
\label{pi(a0)}  
\end{eqnarray}
where $A^\alpha, B^{\alpha\beta}$ and $C^{\alpha\beta} \equiv - \frac{1}{2} \tilde g^{00} g^{\alpha\beta}$ 
have no $\dot g_{\mu\nu}$ and $B^{\alpha\beta} \partial_\beta\phi$ does not include $\dot \phi$. Then, we find 
that the CCR (\ref{pi(a0)-g}) produces:
\begin{eqnarray}
[ g_{\mu\nu}, b_\rho^\prime ] = - i \tilde f ( \delta_\mu^0 g_{\rho\nu} + \delta_\nu^0 g_{\rho\mu} ) \delta^3.
\label{g-b}  
\end{eqnarray}
From this ETCR, it is easy to derive the following two ETCRs:
\begin{eqnarray}
&{}& [ g^{\mu\nu}, b_\rho^\prime ] = i \tilde f ( g^{0\mu} \delta^\nu_\rho + g^{0\nu} \delta^\mu_\rho ) \delta^3,
\nonumber\\
&{}& [ \tilde g^{\mu\nu}, b_\rho^\prime ] = i \tilde f ( \tilde g^{0\mu} \delta^\nu_\rho + \tilde g^{0\nu} \delta^\mu_\rho 
- \tilde g^{\mu\nu} \delta^0_\rho ) \delta^3.
\label{g-b2}  
\end{eqnarray}

Next, we see that the CCR, $[ \pi_\phi, \phi^\prime ] = - i \delta^3$ directly leads to the relation:
\begin{eqnarray}
( \tilde g^{0\mu} g^{0\nu} - \tilde g^{00} g^{\mu\nu} ) [ \dot g_{\mu\nu}, \phi^\prime ] = i \delta^3.
\label{pi-phi-phi}  
\end{eqnarray}
From the symmetry argument, the ETCR, $[ \dot g_{\mu\nu}, \phi^\prime ]$ should have the following expression:
\begin{eqnarray}
[ \dot g_{\mu\nu}, \phi^\prime ] = a_1 ( g_{\mu\nu} + a_2 \delta_\mu^0 \delta_\nu^0 ) \delta^3,
\label{dot-g-phi}  
\end{eqnarray}
where the coefficients $a_1, a_2$ can be determined from Eqs. (\ref{D-eq}) and (\ref{pi-phi-phi}).
Indeed, it turns out that Eq. (\ref{D-eq}) fixes $a_2 = \frac{2}{g^{00}}$ and Eq. (\ref{pi-phi-phi}) does $a_1 = - \frac{i}{3}
\tilde f$, so we have the ETCR:
\begin{eqnarray}
[ \dot g_{\mu\nu}, \phi^\prime ] = - \frac{i}{3} \tilde f \left( g_{\mu\nu} + \frac{2}{g^{00}} \delta_\mu^0 \delta_\nu^0 \right) \delta^3.
\label{dot-g-phi2}  
\end{eqnarray}

We are now ready to evaluate the ETCR between the metric field and its time derivative, i.e., $[ \dot g_{\rho\sigma}, 
g_{\mu\nu}^\prime ]$. To determine this ETCR, we look for conditions imposed on it. 
First of all, let us consider the CCR, $[ \pi_\phi, g^\prime_{\mu\nu} ] = 0$. From Eq. (\ref{CCM}), $\pi_\phi$ is written as
\begin{eqnarray}
\pi_\phi= ( \tilde g^{0 0} g^{\rho\sigma} - \tilde g^{0 \rho} g^{0 \sigma} ) \dot g_{\rho\sigma}
+ ( \tilde g^{0 k} g^{\rho\sigma} - \tilde g^{0 \rho} g^{k \sigma} ) \partial_k g_{\rho\sigma}.
\label{pi-phi}  
\end{eqnarray}
Using this expression, $[ \pi_\phi, g^\prime_{\mu\nu} ] = 0$ provides us with the equation:
\begin{eqnarray}
( g^{0 0} g^{\rho\sigma} - g^{0 \rho} g^{0 \sigma} ) [ \dot g_{\rho\sigma}, g_{\mu\nu}^\prime ] = 0.
\label{pi-phi-g}  
\end{eqnarray}
Moreover, the ${\cal{D}}$-equation (\ref{D-eq}) stemming from the de Donder gauge condition gives rise to
the equation
\begin{eqnarray}
[ {\cal{D}}^{\lambda\rho\sigma} \dot g_{\rho\sigma}, g_{\mu\nu}^\prime ] = 0, 
\label{D-eq2}  
\end{eqnarray}
from which we have the equation for $\lambda= 0$:
\begin{eqnarray}
( g^{0 0} g^{\rho\sigma} - 2 g^{0 \rho} g^{0 \sigma} ) [ \dot g_{\rho\sigma}, g_{\mu\nu}^\prime ] = 0.
\label{D-eq2-2}  
\end{eqnarray}
Eqs. (\ref{pi-phi-g}) and (\ref{D-eq2-2}) together give us the equations:
\begin{eqnarray}
g^{\rho\sigma} [ \dot g_{\rho\sigma}, g_{\mu\nu}^\prime ] = 0, \qquad
g^{0\rho} g^{0 \sigma} [ \dot g_{\rho\sigma}, g_{\mu\nu}^\prime ] = 0.
\label{pi-phi-g-eq}  
\end{eqnarray}

Next, putting $\lambda = i$ in Eq. (\ref{D-eq2}), we have:
\begin{eqnarray}
( g^{0 i} g^{\rho\sigma} - 2 g^{i \rho} g^{0 \sigma} ) [ \dot g_{\rho\sigma}, g_{\mu\nu}^\prime ] = 0.
\label{D-eq2-3}  
\end{eqnarray}
Eqs.  (\ref{pi-phi-g-eq}) and (\ref{D-eq2-3}) are then summarized to two equations:
\begin{eqnarray}
g^{\rho\sigma} [ \dot g_{\rho\sigma}, g_{\mu\nu}^\prime ] = 0, \qquad
g^{0 \sigma} [ \dot g_{\rho\sigma}, g_{\mu\nu}^\prime ] = 0.
\label{pi-phi-g-eq2}  
\end{eqnarray}

Here let us note that $[ \dot g_{\rho\sigma}, g_{\mu\nu}^\prime ]$ has 
a symmetry under the simultaneous exchange 
of $(\mu\nu) \leftrightarrow (\rho\sigma)$ and primed $\leftrightarrow$ unprimed  
in addition to the usual symmetry
$\mu\leftrightarrow \nu$ and $\rho\leftrightarrow \sigma$, because 
the time derivative $\partial_0$ of $[g_{\rho\sigma}, g_{\mu\nu}^\prime ]=0$ leads to 
$[\dot g_{\rho\sigma}, g_{\mu\nu}^\prime ]=[\dot g_{\mu\nu}^\prime, g_{\rho\sigma} ]$.
Assume that this ETCR is proportional to $\delta^3$. Then, we can write down its generic expression
\begin{eqnarray}
[ \dot g_{\rho\sigma}, g_{\mu\nu}^\prime ] &=& \Big\{ c_1 g_{\rho\sigma} g_{\mu\nu} + c_2 ( g_{\rho\mu} g_{\sigma\nu}
+ g_{\rho\nu} g_{\sigma\mu} )
\nonumber\\
&+& \sqrt{-g} \tilde f [ c_3 ( \delta_\rho^0 \delta_\sigma^0 g_{\mu\nu} + \delta_\mu^0 \delta_\nu^0 g_{\rho\sigma} )
+ c_4 ( \delta_\rho^0 \delta_\mu^0 g_{\sigma\nu} + \delta_\rho^0 \delta_\nu^0 g_{\sigma\mu} 
\nonumber\\
&+& \delta_\sigma^0 \delta_\mu^0 g_{\rho\nu} + \delta_\sigma^0 \delta_\nu^0 g_{\rho\mu} ) ] 
+ ( \sqrt{-g} \tilde f )^2 c_5 \delta_\rho^0 \delta_\sigma^0 \delta_\mu^0 \delta_\nu^0 \Big\} \delta^3,  
\label{dot-g-g}  
\end{eqnarray}
where $c_i ( i = 1, \cdots, 5)$ are suitable coefficients to be fixed shortly. By imposing Eq. (\ref{pi-phi-g-eq2}) on (\ref{dot-g-g}),
we find that $c_i ( i = 2, \cdots, 5)$ can be expressed only in terms of $c_1$ as
\begin{eqnarray}
c_2 = - \frac{3}{2} c_1, \quad c_3 = - c_1, \quad c_4 = \frac{3}{2} c_1, \quad c_5 = - 2 c_1.
\label{all-c-i}  
\end{eqnarray}

In order to determine the coefficient $c_1$, we need to calculate the ETCR, $[ \dot g_{kl}, g_{mn}^\prime ]$ explicitly. 
To do so, let us first consider the CCR:
\begin{eqnarray}
[ \pi_g^{kl}, g_{mn}^\prime ] = - i \frac{1}{2} ( \delta_m^k \delta_n^l + \delta_m^l \delta_n^k ) \delta^3
\equiv - i \delta_m^{(k} \delta_n^{l)} \delta^3.
\label{pi-g-kl-g}  
\end{eqnarray}
Next, from Eq. (\ref{CCM}), $\pi_g^{kl}$ can be expanded as
\begin{eqnarray}
\pi_g^{kl} = \hat A^{kl} + \hat B^{kl\rho} b_\rho + \hat C^{klmn} \dot g_{mn} + \hat D^{kl} \dot \phi.
\label{Pi-G-CCM}  
\end{eqnarray}
Here $\hat A^{kl}, \hat B^{kl\rho}, \hat C^{klmn}$ and $\hat D^{kl}$ commute with $g_{mn}$, and 
$\hat C^{klmn}$ and $\hat D^{kl}$ are defined as\footnote{It turns out that the concrete expressions of $\hat A^{kl}$ 
and $\hat B^{kl\rho}$ are irrelevant to the calculation of $[ \dot g_{kl}, g_{mn}^\prime ]$, so we omit to write them down.}
\begin{eqnarray}
\hat C^{klmn} = \frac{1}{2} \sqrt{-g} \left( \frac{1}{2} + \phi \right) K^{klmn}, \qquad
\hat D^{kl} = \tilde g^{00} g^{kl} - \tilde g^{0k} g^{0l},
\label{Pi-G-CCM2}  
\end{eqnarray}
where the definition of $K^{klmn}$ and its property are given by
\begin{eqnarray}
&{}& K^{klmn} = \left|
\begin{array}{rrr}
g^{00} & g^{0l} & g^{0n} \\
g^{k0} & g^{kl} & g^{kn} \\
g^{m0} & g^{ml} & g^{mn} \\
\end{array}
\right|,
\nonumber\\
&{}& K^{klmn} \frac{1}{2} (g^{00})^{-1} ( g_{ij} g_{mn} - g_{im} g_{jn} - g_{in} g_{jm} )
= \frac{1}{2} ( \delta_i^k \delta_j^l + \delta_i^l \delta_j^k )
\equiv \delta_i^{(k} \delta_j^{l)}.
\label{Pi-G-CCM3}  
\end{eqnarray}

From Eq. (\ref{Pi-G-CCM}), we can calculate 
\begin{eqnarray}
[ \dot g_{kl}, g_{mn}^\prime ] &=& \hat C^{-1}_{klpq} \left( [ \pi_g^{pq}, g_{mn}^\prime ] 
-  \hat B^{pq\rho} [ b_\rho, g_{mn}^\prime ] -  \hat D^{pq} [ \dot \phi, g_{mn}^\prime ] \right)
\nonumber\\
&=& \hat C^{-1}_{klpq} \left[ - i \delta^{(p}_m \delta^{q)}_n 
+ i \frac{1}{3} g_{mn} ( \tilde g^{00} g^{pq} - \tilde g^{0p} g^{0q} ) \right] \delta^3,
\label{Pi-G-CCM4}  
\end{eqnarray}
where we have used Eqs. (\ref{g-b}), (\ref{dot-g-phi2})  and (\ref{pi-g-kl-g}). Since we can calculate 
\begin{eqnarray}
\hat C^{-1}_{klpq} = \tilde f \left( \frac{1}{2} + \phi \right)^{-1} ( g_{kl} g_{pq} - g_{kp} g_{lq} - g_{kq} g_{lp} ),
\label{C-inverse}  
\end{eqnarray}
we can eventually arrive at the result:
\begin{eqnarray}
[ \dot g_{kl}, g_{mn}^\prime ] = - i \tilde f \left( \frac{1}{2} + \phi \right)^{-1} 
\left( \frac{2}{3} g_{kl} g_{mn} - g_{km} g_{ln} - g_{kn} g_{lm} \right) \delta^3.
\label{Pi-G-CCM-final}  
\end{eqnarray}
Meanwhile, from Eq. (\ref{dot-g-g}) we have the ETCR:
\begin{eqnarray}
[ \dot g_{kl}, g_{mn}^\prime ] = \left[ c_1 g_{kl} g_{mn} + c_2 ( g_{km} g_{ln} + g_{kn} g_{lm} ) \right] \delta^3
\label{Pi-G-CCM-com}  
\end{eqnarray}
Hence, comparing (\ref{Pi-G-CCM-final}) with (\ref{Pi-G-CCM-com}), we can deduce that
\begin{eqnarray}
c_1 = - \frac{2}{3} i \tilde f \left( \frac{1}{2} + \phi \right)^{-1}, \qquad
c_2 = i \tilde f \left( \frac{1}{2} + \phi \right)^{-1}.
\label{c1-c2}  
\end{eqnarray}
Note that these values satisfy the relation in Eq. (\ref{all-c-i}), $c_2 = - \frac{3}{2} c_1$, which gives us 
a nontrivial verification of our calculation.
In this way, we have succeeded in getting the following ETCR:
\begin{eqnarray}
[ \dot g_{\rho\sigma}, g_{\mu\nu}^\prime ] &=& i \tilde f \left( \frac{1}{2} + \phi \right)^{-1} 
\Big\{ - \frac{2}{3} g_{\rho\sigma} g_{\mu\nu} + g_{\rho\mu} g_{\sigma\nu} + g_{\rho\nu} g_{\sigma\mu} 
\nonumber\\
&+& \sqrt{-g} \tilde f \Big[ \frac{2}{3} ( \delta_\rho^0 \delta_\sigma^0 g_{\mu\nu} 
+ \delta_\mu^0 \delta_\nu^0 g_{\rho\sigma} ) - \delta_\rho^0 \delta_\mu^0 g_{\sigma\nu} 
- \delta_\rho^0 \delta_\nu^0 g_{\sigma\mu}  - \delta_\sigma^0 \delta_\mu^0 g_{\rho\nu} 
- \delta_\sigma^0 \delta_\nu^0 g_{\rho\mu} \Big]
\nonumber\\
&+& ( \sqrt{-g} \tilde f )^2 \frac{4}{3} \delta_\rho^0 \delta_\sigma^0 \delta_\mu^0 \delta_\nu^0 \Big\}
\delta^3.
\label{ETCR-final}  
\end{eqnarray}

It is valuable to compare this ETCR with those of the other gravitational theories.  In general relativity with 
only the Einstein-Hilbert action \cite{Nakanishi, N-O-text}, the ETCR, $[ \dot g_{\rho\sigma}, g_{\mu\nu}^\prime ]$ 
is nonvanishing, but it takes a simpler expression than Eq. (\ref{ETCR-final}). As for conformal gravity \cite{Oda-Ohta}, 
this ETCR is also nonvanishing, but has a rather simple structure compared to that of general relativity and Eq. (\ref{ETCR-final}).  
On the other hand, in the quadratic gravity, which symbolically takes the Lagrangian form, 
$\sqrt{-g} ( R + R^2 + C_{\mu\nu\rho\sigma}^2 )$ with the conventional conformal tensor $C_{\mu\nu\rho\sigma}$, 
it was mentioned that this ETCR identically vanishes \cite{Kimura1, Kimura2, Kimura3}. However, as shown in the
present study, in the $f(R)$ gravity, Eq. (\ref{ETCR-final}) has the nonvanishing and complicated expression. 
Since the $f(R)$ gravity includes the $R + R^2$ gravitational theory in Eq. (\ref{R^2-equiv}) as a special case,
the present result suggests that the $C_{\mu\nu\rho\sigma}^2$ term might play an important role in making this ETCR 
be vanishing in the quadratic gravity. As a final comment, the ETCR (\ref{ETCR-final}) does not reduce to the
corresponding ETCR in general relativity in the limit $\phi \rightarrow 0$. This fact suggests that the canonical
structure for the gravitational sector is quite distinct between $f(R)$ gravity and general relativity.   

Now we would like to present the remaining ETCRs. 
The CCR, $[ \pi_g^{\alpha0}, \phi^\prime ] = 0$ leads to
\begin{eqnarray}
[ b_\rho, \phi^\prime ] = 0.
\label{b-phi}  
\end{eqnarray}
From the CCR, $[ \pi_g^{ij}, \phi^\prime ] = 0$, we can show that 
\begin{eqnarray}
[ \dot \phi, \phi^\prime ] =  i \frac{1}{3} \tilde f \left( \frac{1}{2} + \phi \right) \delta^3.
\label{dot-phi-phi}  
\end{eqnarray}

As for the ETCRs relevant to FP ghosts, we find that the antiCCRs, $\{ \pi_{c \lambda}, c^{\sigma\prime} \}
= \{ \pi_{\bar c}^\sigma, \bar c_\lambda^\prime \} = i \delta_\lambda^\sigma\delta^3$ yield the ETCRs:
\begin{eqnarray}
\{ \dot{\bar c}_\lambda, c^{\prime\sigma} \} = - \{ \dot c^\sigma, \bar c_\lambda^\prime \}
= - \tilde f \delta_\lambda^\sigma\delta^3.
\label{gh-antigh}  
\end{eqnarray}
Moreover, it is easy to see that the CCRs, 
$\{ \pi_g^{\alpha0}, c^{\prime\sigma} \} = \{ \pi_g^{\alpha0}, \bar c_\lambda^\prime \} = 0$ produce:
\begin{eqnarray}
[ b_\rho, c^{\prime\sigma} ] = [ b_\rho, \bar c_\lambda^\prime ] = 0.
\label{b-ghs}  
\end{eqnarray}
The CCRs, $[ g_{\mu\nu}, \pi_{c\lambda}^\prime ] = [ g_{\mu\nu}, \pi_{\bar c}^{\prime\sigma} ] = 0$ directly 
yield: 
\begin{eqnarray}
[ g_{\mu\nu}, \dot c^{\prime\sigma} ] = [ g_{\mu\nu}, \dot{\bar c}_\lambda^\prime ] = 0.
\label{g-dot-ghs}  
\end{eqnarray}
Similarly, the CCRs, $[ \phi, \pi_{c\lambda}^\prime ] = [ \phi, \pi_{\bar c}^{\prime\sigma} ] = 0$ lead to
\begin{eqnarray}
[ \phi, \dot c^{\prime\sigma} ] = [ \phi, \dot{\bar c}_\lambda^\prime ] = 0.
\label{phi-dot-ghs}  
\end{eqnarray}

Next, let us evaluate the type of the ETCRs, $[ \dot \Psi, b_\rho^\prime ]$ where $\Psi$ is a set of
variables $\Psi = \{ g_{\mu\nu}, \phi, c^\sigma, \bar c_\lambda, b_\rho \}$. A little complicated ETCR
is given by 
\begin{eqnarray}
[ \dot g_{\mu\nu}, b_\rho^\prime ] = - i \Bigl\{ \tilde f ( \partial_\rho g_{\mu\nu} + \delta_\mu^0 \dot g_{\rho\nu} 
+ \delta_\nu^0 \dot g_{\rho\mu} ) \delta^3 
+ [ ( \delta_\mu^k - 2 \delta_\mu^0 \tilde f \tilde g^{0 k} ) g_{\rho\nu}
+ (\mu\leftrightarrow \nu) ] \partial_k ( \tilde f \delta^3 ) \Bigr\}. 
\label{dot g-b}  
\end{eqnarray}
This ETCR has been previously derived by using the fact that the translation generator is given by $P_\rho
= \int d^3 x \, \tilde g^{0\lambda} \partial_\lambda b_\rho$. Here we present an alternative derivation based on 
the BRST transformation of the latter equation in Eq. (\ref{g-dot-ghs}). Namely, taking the BRST transformation of
$[ g_{\mu\nu}, \dot{\bar c}_\rho^\prime ] = 0$, we have:
\begin{eqnarray}
[ g_{\mu\nu}, \dot b_\rho^\prime ] = - i [ g_{\mu\nu}, \partial_0 ( c^{\prime\lambda} \partial_\lambda \bar c_\rho^\prime ) ]
- i \{ c^\alpha \partial_\alpha g_{\mu\nu} + \partial_\mu c^\alpha g_{\alpha\nu} + \partial_\nu c^\alpha g_{\mu\alpha},
\dot{\bar c}_\rho^\prime \}.
\label{BRST-g-dot-antigh}  
\end{eqnarray}
The first term on the RHS turns out to be vanishing since
\begin{eqnarray}
- i [ g_{\mu\nu}, \partial_0 ( c^{\prime\lambda} \partial_\lambda \bar c_\rho^\prime ) ]
= - i c^{\prime0} [ g_{\mu\nu}, \ddot{\bar c}_\rho^\prime ]
= 0,
\label{BRST-g-dot-antigh2}  
\end{eqnarray}
where we have used Eq. (\ref{g-dot-ghs}) and $\ddot{\bar c}_\rho = - \tilde f ( 2 \tilde g^{0k} \partial_k \dot{\bar c}_\rho
+ \tilde g^{kl} \partial_k \partial_l \bar c_\rho )$ which is obtained from the field equation 
$g^{\mu\nu} \partial_\mu \partial_\nu \bar c_\rho = 0$. To calculate the second term on the RHS of Eq. (\ref{BRST-g-dot-antigh}),
we need to make use of $[ \dot g_{\mu\nu}, \dot{\bar c}_\rho^\prime ] = 0$, Eq. (\ref{gh-antigh}) and the equation
\begin{eqnarray}
\{ \dot c^\sigma, \dot{\bar c}_\rho^\prime \}
= \delta_\rho^\sigma [ \partial_0 \tilde f + 2 \tilde f \tilde g^{0k} \partial_k ( \tilde f \delta^3 ) ],
\label{Eq-dot c-dot c}  
\end{eqnarray}
which is easily proved by using Eq. (\ref{gh-antigh}) and the field equation for $\bar c_\rho$. 
As a result, we can show that  
\begin{eqnarray}
[ g_{\mu\nu}, \dot b_\rho^\prime ] = i \Bigl\{ [ \tilde f \partial_\rho g_{\mu\nu} - \partial_0 \tilde f ( \delta_\mu^0 g_{\rho\nu} 
+ \delta_\nu^0 g_{\rho\mu} ) ] \delta^3 
+ [ ( \delta_\mu^k - 2 \delta_\mu^0 \tilde f \tilde g^{0 k} ) g_{\rho\nu}
+ (\mu\leftrightarrow \nu) ] \partial_k ( \tilde f \delta^3 ) \Bigr\}.
\label{g-dot b}  
\end{eqnarray}
Then, with the help of Eq. (\ref{g-b}), Eq. (\ref{g-dot b}) produces (\ref{dot g-b}). 
In a similar manner, the BRST transformation makes it possible to derive the following ETCRs:
\begin{eqnarray}
&{}& [ \dot \phi, b_\rho^\prime ] = - i \tilde f \partial_\rho \phi \delta^3,  \qquad
[ \dot{\bar c}_\sigma, b_\rho^\prime ] = - i \tilde f \partial_\rho \bar c_\sigma \delta^3,
\nonumber\\
&{}& [ \dot c^\sigma, b_\rho^\prime ] = - i \tilde f \partial_\rho c^\sigma \delta^3.
\label{dot Phi-b}  
\end{eqnarray}
Note that the second and third equations can be also obtained from the CCRs, 
$[ \pi_{c \lambda}, \pi_g^{\alpha0 \prime} ] = [ \pi_{\bar c}^\sigma, \pi_g^{\alpha0 \prime} ] = 0$.

Finally, let us comment on the ETCR, $[ b_\mu, \dot b_\nu^\prime ]$. First, let us note that
the BRST transformation of $[ b_\mu, \bar c_\nu^\prime ] = 0$ provides us with the ETCR:
\begin{eqnarray}
[ b_\mu, b_\nu^\prime ] = 0.
\label{b-b}  
\end{eqnarray}
Moreover, from two different derivations in Appendix B, we can deduce that
\begin{eqnarray}
[ b_\rho, \dot b_\lambda^\prime ] = i \tilde f ( \partial_\rho b_\lambda+ \partial_\lambda b_\rho) \delta^3.
\label{b-b-rel}  
\end{eqnarray}

\section{Unitarity of physical S-matrix}

In this section, we analyze asymptotic fields under the assumption that all fields have their own
asymptotic fields and there is no bound state. We also assume that all asymptotic fields are
governed by the quadratic part of the quantum Lagrangian apart from possible renormalization.
We will prove the unitarity of the physical S-matrix of $f(R)$ gravity on the basis of the BRST quartet
mechanism \cite{Kugo-Ojima}.

Let us expand the gravitational field $g_{\mu\nu}$ around a flat Minkowski metric $\eta_{\mu\nu}$ as
\begin{eqnarray}
g_{\mu\nu} = \eta_{\mu\nu} + \varphi_{\mu\nu},
\label{Background}  
\end{eqnarray}
where $\varphi_{\mu\nu}$ denotes fluctuations.
For sake of simplicity, we use the same notation for the other asymptotic fields as that for the
interacting fields. Then, up to surface terms the quadratic part of the quantum Lagrangian (\ref{q-Lag}) reads:
\begin{eqnarray}
&{}& {\cal L}_q = \frac{1}{2} \Bigl( \frac{1}{4} \varphi_{\mu\nu} \Box \varphi^{\mu\nu} 
- \frac{1}{4} \varphi \Box \varphi - \frac{1}{2} \varphi^{\mu\nu} \partial_\mu \partial_\rho \varphi_\nu{}^\rho
+ \frac{1}{2} \varphi^{\mu\nu} \partial_\mu \partial_\nu \varphi \Bigr)
\nonumber\\
&{}& - \phi ( \Box \varphi - \partial_\mu \partial_\nu \varphi^{\mu\nu} ) - \frac{1}{4 \alpha} \phi^2
+ \Bigl( \varphi^{\mu\nu} - \frac{1}{2} \eta^{\mu\nu} \varphi \Bigr) \partial_\mu b_\nu
- i \partial_\mu \bar c_\rho \partial^\mu c^\rho. 
\label{Free-Lag}  
\end{eqnarray}
Here note that only the quadratic term in $\Phi(\phi)$, which is represented by $- \frac{1}{4 \alpha} \phi^2$ 
as in Eq. (\ref{R^2-equiv}), makes contribution to this quadratic order equation. To put it differently, only an $R^2$
term in $f(R)$ contributes to kinetic terms while higher-order terms more than the quadratic terms, those are,
$f(R) = c_1 R^3 + c_2 R^4 + \dots$ with $c_1, c_2$ being constants, do to interaction terms since $R$ starts with
linear terms in $\varphi_{\mu\nu}$.
In this section, the spacetime indices $\mu, \nu, \dots$ are raised or lowered by the Minkowski metric $\eta^{\mu\nu}
= \eta_{\mu\nu} = \rm{diag} ( -1, 1, 1, 1)$, and we define $\Box \equiv \eta^{\mu\nu} \partial_\mu \partial_\nu$
and $\varphi \equiv \eta^{\mu\nu} \varphi_{\mu\nu}$.

From this Lagrangian, it is straightforward to derive the following linearized field equations: 
\begin{eqnarray}
&{}& \frac{1}{2} \biggl( \frac{1}{2} \Box \varphi_{\mu\nu} - \frac{1}{2} \eta_{\mu\nu} \Box \varphi 
- \partial_\rho \partial_{(\mu} \varphi_{\nu)}{}^\rho + \frac{1}{2} \partial_\mu \partial_\nu \varphi
+ \frac{1}{2} \eta_{\mu\nu} \partial_\rho \partial_\sigma \varphi^{\rho\sigma} \biggr)
\nonumber\\
&{}& + ( - \eta_{\mu\nu} \Box + \partial_\mu \partial_\nu ) \phi
+ \partial_{(\mu} b_{\nu)} - \frac{1}{2} \eta_{\mu\nu} \partial_\rho b^\rho = 0.
\label{Linear-Eq1}
\\
&{}& \Box \varphi - \partial_\mu \partial_\nu \varphi^{\mu\nu} + \frac{1}{2 \alpha} \phi = 0.
\label{Linear-Eq2}
\\
&{}& \partial^\nu \varphi_{\mu\nu} - \frac{1}{2} \partial_\mu \varphi = 0. 
\label{Linear-Eq3}  
\\
&{}& \Box c^\rho = \Box \bar c_\rho = 0. 
\label{Linear-Eq4}  
\end{eqnarray}
Here, for instance, $\partial_{(\mu} b_{\nu)} \equiv \frac{1}{2} ( \partial_\mu b_\nu + \partial_\nu b_\mu )$.

Now we are ready to simplify the field equations obtained above. 
First, operating $\partial^\mu$ on the linearized Einstein equation (\ref{Linear-Eq1}), we have
\begin{eqnarray}
\Box b_\mu = 0,
\label{L-b-rho-eq}  
\end{eqnarray}
which is a linearized analog of Eq. (\ref{b-field-eq}). This equation can be also obtained by taking 
the linearized BRST transformation $\delta_B^{(L)} \bar c_\mu = i b_\mu$
of $\Box \bar c_\mu = 0$ in Eq. (\ref{Linear-Eq4}).

Next, taking the trace of the linearized Einstein equation (\ref{Linear-Eq1}) and using 
Eq. (\ref{Linear-Eq2}) produces
\begin{eqnarray}
( \Box - m^2 ) \phi + \frac{1}{3} \partial_\rho b^\rho = 0,
\label{L-eq1}  
\end{eqnarray}
where we have defined $m^2 \equiv \frac{1}{12 \alpha} > 0$.
Moreover, acting $\Box$ on this equation and using Eq. (\ref{L-b-rho-eq}), we arrive at
\begin{eqnarray}
\Box ( \Box - m^2 ) \phi = 0.
\label{L-eq2}  
\end{eqnarray}
From Eqs. (\ref{Linear-Eq2}) and (\ref{Linear-Eq3}), it is easy to see that 
\begin{eqnarray}
\phi = - \alpha \Box \varphi.
\label{L-eq3}  
\end{eqnarray}
Then, together with Eq. (\ref{L-eq2}), this equation implies that 
\begin{eqnarray}
\Box^2 ( \Box - m^2 ) \varphi = 0.
\label{L-eq4}  
\end{eqnarray}

Finally, let us focus on the linearized Einstein equation Eq. (\ref{Linear-Eq1}). After some calculations using several equations, 
it turns out that Eq. (\ref{Linear-Eq1}) can be rewritten into a more compact form:
\begin{eqnarray}
&{}& \frac{1}{4} \Box \varphi_{\mu\nu} + \left( \partial_\mu \partial_\nu 
+ \frac{1}{2} \eta_{\mu\nu} \Box \right) \phi + \partial_{(\mu} b_{\nu)} = 0.    
\label{L-Grav-eq}  
\end{eqnarray}
Using Eqs. (\ref{L-b-rho-eq}), (\ref{L-eq2}) and (\ref{L-Grav-eq}), we can obtain the equation for gravitational 
field $\varphi_{\mu\nu}$:
\begin{eqnarray}
\Box^2 \Bigl( \Box - m^2 \Bigr) \varphi_{\mu\nu} = 0.    
\label{L-Grav-eq2}  
\end{eqnarray}
Note that the trace part of this equation reduces to Eq. (\ref{L-eq4}). 

Eq. (\ref{L-Grav-eq2}) implies that there are both massless and massive modes in $\varphi_{\mu\nu}$. In order to disentangle 
these two modes, let us introduce the following two fields: 
\begin{eqnarray}
\tilde \phi &=& \phi - \frac{1}{3 m^2} \partial_\rho b^\rho 
= - \alpha \Box \varphi - \frac{1}{3 m^2} \partial_\rho b^\rho,
\nonumber\\ 
h_{\mu\nu} &=&  \varphi_{\mu\nu} + 2 \eta_{\mu\nu} \phi 
+ 48 \alpha \partial_\mu \partial_\nu \phi - 8 \alpha \eta_{\mu\nu} \partial_\rho b^\rho,
\label{m-scalar-graviton}  
\end{eqnarray}
where Eq. (\ref{L-eq3}) was utilized in the former equation. With these fields, we find that $\tilde \phi$ satisfies the massive
Klein-Gordon equation
\begin{eqnarray}
( \Box - m^2 ) \tilde \phi = 0,
\label{KG-eq}  
\end{eqnarray}
while $h_{\mu\nu}$ obeys both massless dipole ghost equation and the de Donder condition:
\begin{eqnarray}
\Box^2 h_{\mu\nu} = 0, \qquad
\partial^\mu h_{\mu\nu} - \frac{1}{2} \partial_\nu h = 0.
\label{Dipole-eq}  
\end{eqnarray}
In fact, it can be verified shortly by calculating the four-dimensional commutation relation between
$\tilde \phi$ and $h_{\mu\nu}$ that these two fields are independent modes involved in $\varphi_{\mu\nu}$.
Later we will see that the massive scalar field $\tilde \phi$ and two transverse components of $h_{\mu\nu}$,
which is nothing but a massless spin-2 graviton, are physical modes in the theory at hand.

Next, following the standard technique, let us calculate the four-dimensional (anti)commutation 
relations (4D CRs) between asymptotic fields. The point is that the simple pole fields, for instance, 
the Nakanishi-Lautrup field $b_\mu (x)$ obeying $\Box b_\mu = 0$, can be expressed in terms of 
the invariant delta function $D(x)$ as
\begin{eqnarray}
b_\mu (x) = - \int d^3 z D(x-z) \overleftrightarrow{\partial}_0^z b_\mu (z).
\label{D-b1}  
\end{eqnarray}
Here the invariant delta function $D(x)$ for massless simple pole fields and its properties
are described as
\begin{eqnarray}
&{}& D(x) = - \frac{i}{(2 \pi)^3} \int d^4 k \, \epsilon (k^0) \delta (k^2) e^{i k x}, \qquad
\Box D(x) = 0,
\nonumber\\
&{}& D(-x) = - D(x), \qquad D(0, \vec{x}) = 0, \qquad 
\partial_0 D(0, \vec{x}) = - \delta^3 (x), 
\label{D-function}  
\end{eqnarray}
where $\epsilon (k^0) \equiv \frac{k^0}{|k^0|}$. With these properties, it is easy to see that
the right-hand side (RHS) of Eq. (\ref{D-b1}) is independent of $z^0$, and this fact will be
used in evaluating 4D CRs via the ETCRs in what follows.

To illustrate the detail of the calculation, let us evaluate a 4D CR, $[ h_{\mu\nu} (x), b_\rho (y) ]$ explicitly.
Using Eq. (\ref{D-b1}), it can be described as
\begin{eqnarray}
&{}& [ h_{\mu\nu} (x), b_\rho (y) ] 
\nonumber\\
&=& - \int d^3 z D(y-z) \overleftrightarrow{\partial}_0^z [ h_{\mu\nu} (x), b_\rho (z) ]
\nonumber\\
&=& - \int d^3 z \Bigl( D(y-z) [ h_{\mu\nu} (x), \dot b_\rho (z) ] 
- \partial_0^z D(y-z) [ h_{\mu\nu} (x), b_\rho (z) ] \Bigr).
\label{4D-h&b}  
\end{eqnarray}
As mentioned above, since the RHS of Eq. (\ref{D-b1}) is independent of $z^0$, we can put $z^0 = x^0$ 
and use relevant ETCRs:
\begin{eqnarray}
&{}& [ h_{\mu\nu} (x), b_\rho (z) ] = i ( \delta_\mu^0 \eta_{\rho\nu}
+ \delta_\nu^0 \eta_{\rho\mu} ) \delta^3 (x-z),
\nonumber\\
&{}& [ h_{\mu\nu} (x), \dot b_\rho (z) ] = - i ( \delta_\mu^k \eta_{\rho\nu}
+ \delta_\nu^k \eta_{\rho\mu} ) \partial_k \delta^3 (x-z).
\label{4D-h&b2}  
\end{eqnarray}
Substituting Eq. (\ref{4D-h&b2}) into Eq. (\ref{4D-h&b}), we can easily obtain:
\begin{eqnarray}
[ h_{\mu\nu} (x), b_\rho (y) ] 
= - i ( \eta_{\mu\rho} \partial_\nu + \eta_{\nu\rho} \partial_\mu ) D(x-y).
\label{4D-h&b3}  
\end{eqnarray}

In a similar manner, we can calculate the four-dimensional (anti)commutation relations among 
$\tilde \phi, h_{\mu\nu}, b_\mu, c^\mu$ and $\bar c_\mu$. To do that, let us note that since 
$\tilde \phi$ obeys the massive Klein-Gordon equation (\ref{KG-eq}), it can be expressed in terms of the invariant delta function 
$\Delta(x; m^2)$ for massive simple pole fields as
\begin{eqnarray}
\tilde \phi (x) = - \int d^3 z \Delta (x-z; m^2) \overleftrightarrow{\partial}_0^z \tilde \phi (z),
\label{psi-Delta}  
\end{eqnarray}
where $\Delta(x; m^2)$ is defined as
\begin{eqnarray}
&{}& \Delta(x; m^2) = - \frac{i}{(2 \pi)^3} \int d^4 k \, \epsilon (k^0) \delta (k^2 + m^2) e^{i k x}, \quad
(\Box - m^2) \Delta(x; m^2) = 0,
\nonumber\\
&{}& \Delta(-x; m^2) = - \Delta(x; m^2), \quad \Delta(0, \vec{x}; m^2) = 0, 
\nonumber\\
&{}& \partial_0 \Delta(0, \vec{x}; m^2) = - \delta^3 (x),  \qquad
\Delta(x; 0) = D(x). 
\label{Delta-function}  
\end{eqnarray}

As for $h_{\mu\nu}$, since $h_{\mu\nu}$ is a massless dipole ghost field as can be seen 
in Eq. (\ref{Dipole-eq}), it can be described as
\begin{eqnarray}
&{}& h_{\mu\nu} (x) = - \int d^3 z \left[ D(x-z) \overleftrightarrow{\partial}_0^z h_{\mu\nu} (z)
+ E(x-z) \overleftrightarrow{\partial}_0^z \Box h_{\mu\nu} (z) \right]
\nonumber\\
&{}& = - \int d^3 z \left[ D(x-z) \overleftrightarrow{\partial}_0^z h_{\mu\nu} (z)
- 4 E(x-z) \overleftrightarrow{\partial}_0^z \left( \partial_{(\mu} b_{\nu)}
+ \frac{1}{3 m^2} \partial_\mu \partial_\nu \partial_\rho b^\rho \right) (z) \right],
\label{E-varphi}  
\end{eqnarray}
where we have used the equation
\begin{eqnarray}
\Box h_{\mu\nu} = - 4 \left( \partial_{(\mu} b_{\nu)}
+ \frac{1}{3 m^2} \partial_\mu \partial_\nu \partial_\rho b^\rho \right), 
\label{Box-h}  
\end{eqnarray}
which can be derived from Eqs. (\ref{L-eq1}), (\ref{L-Grav-eq}) and (\ref{m-scalar-graviton}), 
and we have introduced the invariant delta function $E(x)$ for massless dipole ghost fields and its properties 
are given by
\begin{eqnarray}
&{}& E(x) = - \frac{i}{(2 \pi)^3} \int d^4 k \, \epsilon (k^0) \delta^\prime (k^2) e^{i k x}, \qquad  
\Box E(x) = D(x),
\nonumber\\
&{}& E(-x) = - E(x), \qquad 
E(0, \vec{x}) = \partial_0 E(0, \vec{x}) = \partial_0^2 E(0, \vec{x}) = 0, 
\nonumber\\ 
&{}& \partial_0^3 E(0, \vec{x}) = \delta^3 (x), \qquad
E(x) = \frac{\partial}{\partial m^2} \Delta(x; m^2)|_{m^2 = 0}. 
\label{E-function}  
\end{eqnarray}
As in Eq. (\ref{D-b1}), we can also show that the RHS of both (\ref{psi-Delta}) and (\ref{E-varphi}) is independent of $z^0$. 

After straightforward calculations, we find the following 4D CRs:
\begin{eqnarray}
&{}& [ \tilde \phi (x), \tilde \phi (y) ] = i \frac{1}{6}  \Delta (x-y; m^2).
\label{4D-CR1}
\\
&{}& [ h_{\mu\nu} (x), h_{\sigma\tau} (y) ] = - 2 i \Bigl[ \eta_{\mu\nu} \eta_{\sigma\tau}
- \eta_{\mu\sigma} \eta_{\nu\tau} - \eta_{\mu\tau} \eta_{\nu\sigma}
+ \frac{2}{3 m^2} ( \eta_{\mu\nu} \partial_\sigma \partial_\tau + \eta_{\sigma\tau} \partial_\mu \partial_\nu )
\nonumber\\
&{}& - \frac{4}{3 m^4} \partial_\mu \partial_\nu \partial_\sigma \partial_\tau \Bigr] D (x-y)
- 2 i \Bigl( \eta_{\mu\sigma} \partial_\nu \partial_\tau + \eta_{\mu\tau} \partial_\nu \partial_\sigma
+ \eta_{\nu\sigma} \partial_\mu \partial_\tau  + \eta_{\nu\tau} \partial_\mu \partial_\sigma 
\nonumber\\
&{}& + \frac{4}{3 m^2} \partial_\mu \partial_\nu \partial_\sigma \partial_\tau \Bigr) E (x-y).
\label{4D-CR2}
\\
&{}& [ h_{\mu\nu} (x), \tilde \phi (y) ] = 0.
\label{4D-CR3}
\\
&{}& [ \tilde \phi (x), b_\rho (y) ] = 0.
\label{4D-CR4}
\\
&{}& [ h_{\mu\nu} (x), b_\rho (y) ] = - i ( \eta_{\mu\rho} \partial_\nu + \eta_{\nu\rho} \partial_\mu ) D(x-y).
\label{4D-CR5}
\\
&{}& \{ c^\mu (x), \bar c_\nu (y) \} = \delta_\nu^\mu D(x-y). 
\label{4D-CR6}
\end{eqnarray}
Here let us note that Eq. (\ref{4D-CR3}) implies that two fields $h_{\mu\nu}$ and $\tilde \phi$ are independent fields.

As usual, the physical Hilbert space $|\rm{phys} \rangle$ is defined by the Kugo-Ojima subsidiary 
conditions \cite{Kugo-Ojima}
\begin{eqnarray}
\rm{Q_B} |\rm{phys} \rangle = 0,
\label{Phys-Hilbert}  
\end{eqnarray}
where $\rm{Q_B}$ is the BRST charge associated with the general coordinate transformation (GCT). 

Now we would like to discuss the issue of the unitarity of the theory at hand. To do that, it is
convenient to perform the Fourier transformation of Eqs. (\ref{4D-CR1})-(\ref{4D-CR6}).
However, for the dipole field we cannot use the three-dimensional Fourier expansion to define 
the creation and annihilation operators. We therefore make use of the four-dimensional 
Fourier expansion \cite{N-O-text}:\footnote{The Fourier transform of a field is denoted 
by the same field except for the argument $x$ or $p$, for simplicity.}
\begin{eqnarray}
\varphi_{\mu\nu} (x) = \frac{1}{(2 \pi)^{\frac{3}{2}}} \int d^4 p \, \theta (p^0) [ \varphi_{\mu\nu} (p) e^{i p x}
+ \varphi_{\mu\nu}^\dagger (p) e^{- i p x} ],
\label{FT-varphi}  
\end{eqnarray}
where $\theta (p^0)$ is the unit step function. For any simple pole fields, we adopt the same Fourier expansion,
for instance, 
\begin{eqnarray}
\beta_\mu (x) = \frac{1}{(2 \pi)^{\frac{3}{2}}} \int d^4 p \, \theta (p^0) [ \beta_\mu (p) e^{i p x}
+ \beta_\mu^\dagger (p) e^{- i p x} ].
\label{FT-beta}  
\end{eqnarray}
Incidentally, for a generic, massless simple pole field $\Phi$, the three-dimensional Fourier expansion is defined as
\begin{eqnarray}
\Phi (x) = \frac{1}{(2 \pi)^{\frac{3}{2}}} \int d^3 p \, \frac{1}{\sqrt{2 |\vec{p}|}}  
[ \Phi (\vec{p}) e^{- i |\vec{p}| x^0 + i \vec{p} \cdot \vec{x} } 
+ \Phi^\dagger (\vec{p}) e^{ i |\vec{p}| x^0 - i \vec{p} \cdot \vec{x} } ],
\label{3D-FT}  
\end{eqnarray}
whereas the four-dimensional Fourier expansion reads:
\begin{eqnarray}
\Phi (x) = \frac{1}{(2 \pi)^{\frac{3}{2}}} \int d^4 p \, \theta (p^0) [ \Phi (p) e^{i p x}
+ \Phi^\dagger (p) (p) e^{- i p x} ].
\label{4D-FT}  
\end{eqnarray}
Thus, the annihilation operator $\Phi (p)$ in the four-dimensional Fourier expansion has a connection with 
the annihilation operator $\Phi (\vec{p})$ in the three-dimensional Fourier expansion like\footnote{For a massive
field, $\delta(p^2)$ and $|\vec{p}|$ are simply replaced with $\delta(p^2 + m^2)$ and $\sqrt{\vec{p}^2 + m^2}$,
respectively.}
\begin{eqnarray}
\Phi (p) = \theta (p^0) \delta (p^2) \sqrt{2 |\vec{p}|} \Phi (\vec{p}).
\label{3D-4D}  
\end{eqnarray}
Note that the Fourier transform $\Phi (p)$ takes the form for a massless simple pole field:
\begin{eqnarray}
\Phi (p) = \frac{i}{(2 \pi)^{\frac{3}{2}}} \theta (p^0) \delta(p^2) \int d^3 z \, e^{-ipz}
\overleftrightarrow{\partial}_0^z \Phi (z).
\label{Phi-p}  
\end{eqnarray}
Based on these Fourier expansions, we can calculate the Fourier transform of Eqs. (\ref{4D-CR1})-(\ref{4D-CR6}):
\begin{eqnarray}
&{}& [ \tilde \phi (p), \tilde \phi^\dagger (q) ] = \frac{1}{6} \theta (p^0) \delta(p^2 + m^2) \delta^4 (p-q).
\label{FT-4D-CR1}
\\
&{}& [ h_{\mu\nu} (p), h_{\sigma\tau}^\dagger (q) ] = \frac{1}{2} \theta (p^0) \delta^4 (p-q)
\Big\{ \delta(p^2) \Big[ \eta_{\mu\sigma} \eta_{\nu\tau} + \eta_{\mu\tau} \eta_{\nu\sigma} - \eta_{\mu\nu} \eta_{\sigma\tau} 
\nonumber\\
&{}& + \frac{2}{3 m^2} ( \eta_{\mu\nu} p_\sigma p_\tau + \eta_{\sigma\tau} p_\mu p_\nu )
+ \frac{4}{3 m^4} p_\mu p_\nu p_\sigma p_\tau \Big]
\nonumber\\
&{}& + 3  \delta^\prime (p^2) \Big( \eta_{\mu\sigma} p_\nu p_\tau + \eta_{\nu\sigma} p_\mu p_\tau +
\eta_{\mu\tau} p_\nu p_\sigma + \eta_{\nu\tau} p_\mu p_\sigma - \frac{4}{3 m^2} p_\mu p_\nu p_\sigma p_\tau \Big) \Big\}.
\label{FT-4D-CR2}
\\
&{}& [ h_{\mu\nu} (p), b_\rho^\dagger (q) ] = - i \frac{1}{2} ( \eta_{\mu\rho} p_\nu + \eta_{\nu\rho} p_\mu ) 
\theta (p^0) \delta(p^2) \delta^4 (p-q). 
\label{FT-4D-CR3}
\\
&{}& \{ c^\mu (p), \bar c_\nu^\dagger (q) \} = - i \delta^\mu_\nu \theta (p^0) \delta(p^2) \delta^4 (p-q). 
\label{FT-4D-CR4}  
\end{eqnarray}

Next, let us turn our attention to the linearized field equations. In the Fourier transformation, 
the de Donder condition in Eq. (\ref{Dipole-eq}) takes the form:
\begin{eqnarray}
p^\nu h_{\mu\nu} - \frac{1}{2} p_\mu h = 0.
\label{FT-Linear-Eq3}
\end{eqnarray}
Since this equation gives us four independent relations in ten components of $h_{\mu\nu} (p)$,
the number of the independent components of $h_{\mu\nu} (p)$ is six. To deal with
the six independent components, it is convenient to take a specific Lorentz 
frame which is defined by $p^\mu = ( p, 0, 0, p )$ with $p > 0$, and choose the six components as follows:
\begin{eqnarray}
&{}& h_1 (p) = h_{11} (p),  \qquad
h_2 (p) = h_{12} (p),  \qquad
\omega_0 (p) = \frac{1}{2 p} [ h_{00} (p) - h_{11} (p) ],   
\nonumber\\
&{}& \omega_I (p) = \frac{1}{p} h_{0I} (p),  \qquad
\omega_3 (p) = - \frac{1}{2 p} [ h_{11} (p) + h_{33} (p) ], 
\label{Lorentz}  
\end{eqnarray}
where the index $I$ takes the transverse components $I = 1, 2$. The other four components
are expressible by the above six ones. 

In this respect, it is worthwhile to consider the BRST transformation for asymptotic fields.
Since the BRST transformation for the Fourier transform of the asymptotic fields is given by
\begin{eqnarray}
&{}& \delta_B \tilde \phi (p) = 0,  \qquad 
\delta_B h_{\mu\nu} (p) = - i [ p_\mu c_\nu (p) + p_\nu c_\mu (p) ], \qquad
\delta_B b_\mu (p) = 0, 
\nonumber\\
&{}& \delta_B \bar c_\mu (p) = i b_\mu (p), \qquad
\delta_B c^\mu (p) = 0, 
\label{Q_B-FT}  
\end{eqnarray}
the BRST transformation in terms of the components in (\ref{Lorentz}) takes the form:
\begin{eqnarray}
&{}& \delta_B \tilde \phi (p) = 0,  \qquad
\delta_B h_I (p) = 0, \qquad
\delta_B \omega_\mu (p) = i c_\mu (p),
\nonumber\\
&{}& \delta_B \bar c_\mu (p) = i b_\mu (p),  \qquad
\delta_B c^\mu (p) = \delta_B b_\mu (p) = 0.
\label{Q_B-Comp}  
\end{eqnarray}
This BRST transformation implies that $\tilde \phi (p)$ and $h_I (p)$
could be physical observables while a set of fields, $\{ \omega_\mu (p), b_\mu (p), 
c_\mu (p), \bar c_\mu (p) \}$ might belong to a BRST quartet, which are dropped
from the physical state by the Kugo-Ojima subsidiary condition, $Q_B | \rm{phys} \rangle = 0$ \cite{Kugo-Ojima}.  
However, note that $b_\mu (p), c_\mu (p)$ and $\bar c_\mu (p)$ are
simple pole fields obeying $p^2 b_\mu (p) = p^2 c_\mu (p) = p^2 \bar c_\mu (p) = 0$,
whereas $h_{\mu\nu} (p)$ is a dipole field satisfying $( p^2 )^2 h_{\mu\nu} (p) = 0$, 
so that a naive Kugo-Ojima's quartet mechanism does not work in a straightforward manner. 

To clarify the BRST quartet mechanism, let us present 4D CRs in terms of the components in (\ref{Lorentz}). 
From Eqs. (\ref{FT-4D-CR1})-(\ref{FT-4D-CR4}) and the definition (\ref{Lorentz}), it is straightforward to 
derive the following 4D CRs:
\begin{eqnarray}
&{}& [ \tilde \phi (p), \tilde \phi^\dagger (q) ] = \frac{1}{6} \theta (p^0) \delta(p^2 + m^2) \delta^4 (p-q).
\label{Lor-4D-CR1}
\\
&{}& [ h_I (p), h_J^\dagger (q) ] = \frac{1}{2} \delta_{IJ} \theta (p^0) \delta(p^2) \delta^4 (p-q).
\label{Lor-4D-CR2}
\\
&{}& [ h_I (p), \omega_\mu^\dagger (q) ] = [ h_I (p), b_\mu^\dagger (q) ]
= [ b_\mu (p), b_\nu^\dagger (q) ] = 0. 
\label{Lor-4D-CR3}
\\
&{}& [ \omega_\mu (p), b_\nu^\dagger (q) ] = i \frac{1}{2} \eta_{\mu\nu} \theta (p^0) 
\delta(p^2) \delta^4 (p-q). 
\label{Lor-4D-CR4}
\\
&{}& \{ c^\mu (p), \bar c_\nu^\dagger (q) \} = - i \delta_\nu^\mu \theta (p^0) 
\delta(p^2) \delta^4 (p-q). 
\label{Lor-4D-CR5}
\end{eqnarray}
In addition to them, we have a bit complicated expression for $[ \omega_\mu (p), \omega_\nu^\dagger (q) ]$
because $h_{\mu\nu} (p)$ is a dipole field, but luckily enough this expression is unnecessary
for our purpose \cite{Kugo-Ojima}. 
It is known how to take out a simple pole field from a dipole field, which amounts to
using an operator defined by \cite{Kugo-Ojima} 
\begin{eqnarray}
{\cal D}_p = \frac{1}{2 |\vec{p}|^2} p_0 \frac{\partial}{\partial p_0} + c, 
\label{D-ope}
\end{eqnarray}
where $c$ is a constant. Using this operator, we can define a simple pole field
$\hat h_{\mu\nu} (p)$ from the dipole field $h_{\mu\nu} (p)$, which
obeys $(p^2)^2 h_{\mu\nu} (p) = 0$, as
\begin{eqnarray}
\hat h_{\mu\nu} (p) &\equiv& h_{\mu\nu} (p) - {\cal D}_p ( p^2 h_{\mu\nu} (p) )
\nonumber\\
&=& h_{\mu\nu} (p) - 4 i {\cal D}_p \left( p_{(\mu} b_{\nu)} (p) - \frac{1}{3 m^2} 
p_\mu p_\nu p_\rho b^\rho (p) \right),
\label{Simple-field}
\end{eqnarray}
where in the last equality we have used the Fourier transform of the linearized field equation (\ref{Box-h}).
It is then easy to verify the equation:
\begin{eqnarray}
p^2 \hat h_{\mu\nu} (p) = 0.
\label{Simple-field2}
\end{eqnarray}
Then, we replace $h_{\mu\nu}$ of $\omega_\mu$ with $\hat h_{\mu\nu}$ in Eq. (\ref{Lorentz}),
and we redefine $\omega_\mu$ by $\hat \omega_\mu$ as
\begin{eqnarray}
&{}& \hat \omega_0 (p) = \frac{1}{2 p} [ \hat h_{00} (p) - \hat h_{11} (p) ],  \qquad
\hat \omega_I (p) = \frac{1}{p} \hat h_{0I} (p),  
\nonumber\\
&{}& \hat \omega_3 (p) = - \frac{1}{2 p} [ \hat h_{11} (p) + \hat h_{33} (p) ].
\label{hat-omega}  
\end{eqnarray}
The key point is that with this redefinition from $\omega_\mu$ to $\hat \omega_\mu$,
the BRST transformation and the 4D CRs remain unchanged owing to
$\delta_B b_\mu = 0$ and $[ b_\mu (p), b^\dagger_\nu (q) ] = [ h_I (p), b^\dagger_\mu (q) ] = 0$, those are,
\begin{eqnarray}
&{}& \delta_B  \hat \omega_\mu (p) = i c_\mu (p),  \quad
[ \hat \omega_\mu (p), b^\dagger_\nu (q) ] = [ \omega_\mu (p), b^\dagger_\nu (q) ],
\nonumber\\
&{}& [ h_I (p), \hat \omega^\dagger_\mu (q) ] = [ h_I (p), \omega^\dagger_\mu (q) ].
\label{point}  
\end{eqnarray}

Since it turns out that all the fields, $\{ \tilde \phi, h_I, \hat \omega_\mu, b_\mu,
c_\mu, \bar c_\mu \}$ are simple pole fields,\footnote{Without the redefinition,
$h_I (p)$ is already a simple pole field as can be seen in Eq. (\ref{Lor-4D-CR2}).}
we can obtain the standard creation and annihilation operators in the three-dimensional 
Fourier expansion from those in the four-dimensional one through the relation (\ref{3D-4D}). 
As a result, the three-dimensional (anti)commutation relations, which are denoted as 
$[ \Phi (\vec{p}), \Phi^\dagger (\vec{q}) \}$ with
$\Phi (\vec{p}) \equiv \{ \tilde \phi (\vec{p}), h_I (\vec{p}), \hat \omega_\mu (\vec{p}), b_\mu (\vec{p}), 
c_\mu (\vec{p}), \bar c_\mu (\vec{p}) \}$, are given by\footnote{The bracket $[ A, B \}$ 
is the graded commutation relation denoting either commutator or anticommutator, according to 
the Grassmann-even or odd character of $A$ and $B$, i.e., $[ A, B \} = A B - (-)^{|A| |B|} B A$.}  
\begin{eqnarray}
[ \Phi (\vec{p}), \Phi^\dagger (\vec{q}) \} &=&
\left(
\begin{array} {cc|cc|cc|cc}
    & \frac{1}{6}&                 &    &    &    &     \\ 
\hline
    &       &     & \frac{1}{2} \delta_{IJ}     &     &    &    &    \\ 
\hline
    &        &      &    & [ \hat \omega_\mu (\vec{p}), \hat \omega_\nu^\dagger (\vec{q}) ]  & i \frac{1}{2} \eta_{\mu\nu}  \\ 
    &        &      &    &  -i \frac{1}{2}  \eta_{\mu\nu}   &  0       &              \\
\hline   
    &        &     &          &       &    &  & - i \eta_{\mu\nu} \\  
    &        &     &          &      & & -i \eta_{\mu\nu} &            \\
\end{array}
\right) 
\nonumber\\
&{}& \times \delta ( \vec{p} - \vec{q} )_.
\label{G-3D-CRs}  
\end{eqnarray}
The (anti)commuatation relations (\ref{G-3D-CRs}) have in essence the same structure as those 
of the Yang-Mills theory \cite{Kugo-Ojima}. In particular, the positive coefficients $\frac{1}{6}$ and
$\frac{1}{2} \delta_{IJ}$ mean that $\tilde \phi$ and $h_I$ have positive norm.
Hence, we find that $\tilde \phi$ and $h_I$, which correspond to
transverse gravitons, are physical observables of positive norm while a set of fields $\{ \hat \omega_\mu, b_\mu, c_\mu, 
\bar c_\mu \}$ belongs to a BRST quartet. This quartet appears in the physical subspace only 
as zero norm states by the Kugo-Ojima subsidiary conditions (\ref{Phys-Hilbert}).
It is worthwhile to stress that both the massive scalar and the massless graviton of positive 
norm appear in the physical Hilbert space so the physical S-matrix is unitary in the present theory.

\section{Conclusions}

In this paper, we have performed the manifestly covariant quantization of $f(R)$ gravity in the de Donder gauge
condition, clarified its global symmetry and calculated the whole equal-time (anti)commutation relation (ETCR) among fundamental 
variables on the basis of BRST formalism. Via this quantization procedure, we have clearly demonstrated that physical modes 
are a massive scalar and the massless graviton, both of which have positive norm, and the other dynamical
degrees of freedom such as FP ghosts emerge in the physical subspace defined by the physical condition,
$Q_B | \rm{phys} \rangle = 0$ only as zero norm states. Thus, we can definitely conclude that the physical S-matrix
is manifestly unitary in the $f(R)$ gravity although it is not renormalizable perturbatively.      

One of motivations behind the present study is to understand the canonical structure of the gravitational sector
in gravitational theories possessing higher-derivative terms. As shown in the present paper, the ETCR between the metric
tensor and its time derivative, $[ \dot g_{\rho\sigma}, g_{\mu\nu}^\prime ]$ takes a nontrivial expression 
while the one in quadratic gravity was mentioned to be vanishing \cite{Kimura1, Kimura2, Kimura3}. Though the physical 
meaning of this vanishing ETCR is not understood yet, it is certain that this triviality comes from the squared term 
of conformal tensor, $\sqrt{-g} \, C_{\mu\nu\rho\sigma}^2$ in the action of the quadratic gravity as shown in the present article 
(In this context, we consider a specific case, $f(R) = \alpha R^2$).   Indeed, in our previous work \cite{Oda-Ohta} 
the BRST formalism of conformal gravity has been investigated where it has been found that the ETCR between 
the metric and its time derivative has a very simple expression though it is not vanishing.    

Related to this fact, it is useful to recall that the quadratic gravity is renormalizable by the power counting, 
but it is not unitary because of the existence of massive ghost \cite{Stelle1}. This breakdown of the unitarity arises from 
the squared term of conformal tensor since $f(R)$ gravity is manifestly unitary as seen in this article. We are somewhat 
in a tantalizing situation. We need the squared term of conformal tensor for ensuring the renormalizability of a theory, 
but this term pays a massive price for the unitarity issue.  In any case, in order to obtain a consistent quantum gravity
which is renormalizable and unitary at the same time, it seems to be necessary to have a deep understanding of the squared term 
of conformal tensor.


\appendix
\addcontentsline{toc}{section}{Appendix~\ref{app:scripts}: Training Scripts}
\section*{Appendix}
\label{app:scripts}
\renewcommand{\theequation}{A.\arabic{equation}}
\setcounter{equation}{0}

\section{Two derivations of Eq. (\ref{b-field-eq})}

In this appendix, we present two different derivations of field equation for $b_\mu$ field since the $b_\mu$
plays an important role as the generator of a global symmetry in the formalism at hand.  In the first derivation 
we make use of field equations while in the second derivation we rely on the BRST transformation.  

Let us present a derivation based on field equations first. We take a covariant derivative of the Einstein equation, 
i.e., the first equation in Eq. (\ref{Field-eq}).  The result reads
\begin{eqnarray}
\nabla^\mu \phi \, G_{\mu\nu} - R_{\mu\nu} \nabla^\mu \phi - \frac{1}{2} \nabla_\nu \Phi 
- \frac{1}{2} \nabla^\mu \left( E_{\mu\nu} - \frac{1}{2} g_{\mu\nu} E \right) = 0,
\label{Cov-Einstein}  
\end{eqnarray}
where we have used the Bianchi identity, $\nabla^\mu G_{\mu\nu} = 0$ and a relation,
$\nabla^\mu ( \nabla_\mu \nabla_\nu - g_{\mu\nu} \Box ) \phi = R_{\mu\nu} \nabla^\mu \phi$.
Then, using the field equation for $\phi$ in Eq. (\ref{Field-eq}), i.e., $R + \Phi^\prime (\phi) = 0$, the
three terms except for the last term on the LHS of the above equation identically vanish. 
Thus, we can obtain:
\begin{eqnarray}
\nabla_\nu \left( E^\nu{}_\mu - \frac{1}{2} \delta_\mu^\nu E \right) = 0.
\label{Cov-E}  
\end{eqnarray}

Generally for a symmetric tensor $S^{\mu\nu}$, we have a formula:
\begin{equation}
\nabla_\nu S^\nu{}_\mu= \frac{1}{\sqrt{-g}} \partial_\nu( \sqrt{-g} \, S^\nu{}_\mu) 
+ \frac{1}{2} S_{\nu\rho} \partial_\mu g^{\nu\rho}.
\label{S}  
\end{equation}
Using this, we can show an equality
\begin{eqnarray}
\nabla_\nu E^\nu{}_\mu = 
\frac{1}{\sqrt{-g}} \partial_\nu \tilde g^{\nu\rho}\cdot E_{\rho\mu} + \partial^2 b_\mu 
+ i \left ( \partial^2\bar c_\lambda\cdot \partial_\mu c^\lambda 
+ \partial_\mu\bar c_\lambda\cdot \partial^2c^\lambda \right)
+ \frac{1}{2} \partial_\mu E,
\label{E-id}  
\end{eqnarray}
where $\partial^2 \equiv g^{\mu\nu} \partial_\mu \partial_\nu$.
Then the equation (\ref{Cov-E}), with the help of the other field equations in Eq. (\ref{Field-eq}), 
is seen to lead to the field equation (\ref{b-field-eq}) for the $b_\mu$ field. 

Next, let us prove the same equation (\ref{b-field-eq}) by using the BRST transformation.
This derivation is a bit simpler compared to the above one.
Let us start with the field equation $g^{\mu\nu} \partial_\mu \partial_\nu \bar c_\rho = 0$.
Taking the BRST transformation of this field equation produces:
\begin{eqnarray}
( - c^\lambda \partial_\nu g^{\mu\nu} + g^{\mu\lambda} \partial_\lambda c^\nu
+ g^{\nu\lambda} \partial_\lambda c^\mu ) \partial_\mu \partial_\nu \bar c_\rho 
+ i g^{\mu\nu} \partial_\mu \partial_\nu B_\rho = 0.
\label{BRST-c-field}  
\end{eqnarray}
Then, rewriting $B_\rho$ in terms of $b_\rho$ via Eq. (\ref{b-field}) in this equation, and using the field equations
for FP (anti)ghosts,  we can easily arrive at the desired equation (\ref{b-field-eq}).

\renewcommand{\theequation}{B.\arabic{equation}}
\setcounter{equation}{0}

\section{Two derivations of $[ b_\mu, \dot b_\nu^\prime ]$}

In this appendix, we present two different derivations of $[ b_\mu, \dot b_\nu^\prime ]$ in Eq. (\ref{b-b-rel}), 
one of which is based on field equations and the other is so on BRST transformation.

In the first derivation, let us first solve $\dot b_\rho$ from Eq. (\ref{E}):
\begin{eqnarray}
\dot b_0 &=& \frac{1}{2} E_{00} - i \dot{\bar c}_\rho\dot c^\rho,
\nonumber\\
\dot b_k &=& E_{0 k} - \partial_k b_0 - i ( \dot{\bar c}_\rho\partial_k c^\rho+ \partial_k \bar c_\rho\dot c^\rho).
\label{dot b}  
\end{eqnarray}
Next, let us rewrite the Einstein equation in Eq. (\ref{Field-eq}) into the form: 
\begin{eqnarray}
E^\mu{}_\nu- \frac{1}{2} \delta^\mu_\nu E = 2 \left( \frac{1}{2} + \phi \right) G^\mu{}_\nu
- 2 ( \nabla^\mu\nabla_\nu- \delta^\mu_\nu\Box ) \phi - \delta^\mu_\nu \Phi(\phi).
\label{Eins-eq}  
\end{eqnarray}
Since we find that $R_{00} \sim- \frac{1}{2} g^{ij} \ddot g_{ij}, R_{0i} \sim\frac{1}{2} g^{0j} \ddot g_{ij}$ and 
$R_{ij} \sim- \frac{1}{2} g^{00} \ddot g_{ij}$, it turns out that $G^0{}_\nu= R^0{}_\nu- \frac{1}{2} \delta^0_\nu R$ contains 
no $\ddot g_{ij}$. Then, we can rewrite $\dot b_\rho$ of Eq. (\ref{dot b}) as
\begin{eqnarray}
\dot b_0 &=& \frac{2}{g^{00}} \Bigl[ \left( \frac{1}{2} + \phi \right) G^0{}_0 - ( \nabla^0 \nabla_0 - \Box ) \phi 
- \frac{1}{2} \Phi(\phi) + \frac{1}{4} g^{ij} E_{ij} \Bigr] - i \dot{\bar c}_\rho\dot c^\rho,
\nonumber\\
\dot b_k &=& \frac{2}{g^{00}} \Bigl[ \left( \frac{1}{2} + \phi \right) R^0{}_k - \nabla^0 \nabla_k \phi- \frac{1}{2} g^{0j} E_{kj} \Bigr] 
- \partial_k b_0 - i ( \dot{\bar c}_\rho\partial_k c^\rho+ \partial_k \bar c_\rho\dot c^\rho).
\label{dot b-2}  
\end{eqnarray}

After some calculations, we can show that 
\begin{eqnarray}
[ G^0{}_\nu, b_\rho^\prime ] = i \tilde f ( \delta^0_\rho R^0{}_\nu- \delta^0_\nu R^0{}_\rho) \delta^3.
\label{G-b}  
\end{eqnarray}
Furthermore, from Eqs. (\ref{b-phi}) and (\ref{dot Phi-b}), we can also show that 
\begin{eqnarray}
[ \nabla_k \nabla_\mu\phi, b_\rho^\prime ] = - i \tilde f \delta^0_\mu\nabla_k \nabla_\rho\phi\delta^3.
\label{nabla-phi-b}  
\end{eqnarray}
To derive this equation, we need to use the equation
\begin{eqnarray}
[ \Gamma^\rho_{\mu\nu}, b_\lambda^\prime ] = i \tilde f ( \delta_\lambda^\rho \Gamma_{\mu\nu}^0 
- \delta_\mu^0 \Gamma_{\lambda\nu}^\rho - \delta_\nu^0 \Gamma_{\mu\lambda}^\rho ) \delta^3
+ i \delta_\lambda^\rho ( 2 \delta^0_\mu \delta^0_\nu \tilde f \tilde g^{0k} 
- \delta^0_\mu \delta^k_\nu - \delta^0_\nu \delta^k_\mu ) \partial_k ( \tilde f \delta^3 ),
\label{Gamma-b}  
\end{eqnarray}
where Eq. (\ref{dot g-b}) was used. Then, it is straightforward to prove the ETCR in Eq. (\ref{b-b-rel}).

In the second derivation of Eq. (\ref{b-b-rel}), we make use of the BRST transformation.
To do that, let us begin by the ETCR, $[ \pi_{c\mu}, b_\nu^\prime ] = 0$, which can be easily proved. 
Taking the BRST transformation of this equation, it is straightforward to obtain the following equation:
\begin{eqnarray}
[ b_\mu, \dot b_\nu^\prime ] = i \tilde f ( \partial_\mu b_\nu + \partial_\nu b_\mu ) \delta^3 
+ A_{\mu\nu} + B_{\mu\nu} + C_{\mu\nu}.
\label{BRST-pi-b}  
\end{eqnarray}
Here $A_{\mu\nu}, B_{\mu\nu}$ and  $C_{\mu\nu}$ are defined as
\begin{eqnarray}
A_{\mu\nu} &=& - i \tilde f [ ( \tilde g^{\rho\sigma} \nabla_\sigma c^0 + \tilde g^{0\sigma} \nabla_\sigma c^\rho
- \tilde g^{0\rho} \nabla_\lambda c^\lambda )  \partial_\rho \bar c_\mu,  b_\nu^\prime ],
\nonumber\\ 
B_{\mu\nu} &=& i \tilde f [ \tilde g^{0\rho} \partial_\rho ( c^\lambda \partial_\lambda \bar c_\mu ),  b_\nu^\prime ],
\nonumber\\ 
C_{\mu\nu} &=& - \tilde f c^{\prime\lambda} [ \pi_{c\mu},  \partial_\lambda b_\nu^\prime ].
\label{ABC}  
\end{eqnarray}
For the evaluation of them, we utilize the following ETCRs:
\begin{eqnarray}
&{}& [ \nabla_\sigma c^\rho, b_\nu^\prime ] = i \tilde f ( - \delta_\sigma^0 \nabla_\nu c^\rho
+ \delta_\nu^\rho \Gamma^0_{\sigma\lambda} c^\lambda - \Gamma^\rho_{\sigma\nu} c^0 ) \delta^3
+ i \delta_\nu^\rho ( 2 \delta_\sigma^0 c^0 \tilde f \tilde g^{0k} - \delta_\sigma^0 c^k
- \delta_\sigma^k c^0 ) \partial_k ( \tilde f \delta^3 ),
\nonumber\\ 
&{}& [ \ddot{\bar c}_\mu, b_\nu^\prime ] = - 2 i \tilde f [ \partial_\nu \dot{\bar c}_\mu \delta^3 
- \tilde g^{0k} \partial_\nu \bar c_\mu \partial_k ( \tilde f \delta^3 ) ].
\label{ABC-formula}  
\end{eqnarray}
After some calculations, we find that they are concretely given by
\begin{eqnarray}
A_{\mu\nu} &=& \tilde f^2 [ - \tilde g^{\rho\sigma} ( \delta_\nu^0 \partial_\sigma c^0 + \Gamma^0_{\sigma\nu} c^0 )
- \tilde g^{0\sigma} ( \delta_\nu^\rho \partial_\sigma c^0 + \Gamma^\rho_{\sigma\nu} c^0 )
+ \tilde g^{\rho0} ( \partial_\nu c^0 + \Gamma^\lambda_{\nu\lambda} c^0 ) ] \partial_\rho \bar c_\mu \delta^3
\nonumber\\
&+& \tilde f [ - \tilde g^{\rho\sigma} \delta_\nu^0 \delta_\sigma^k c^0 + \tilde g^{0\sigma} \delta_\nu^\rho ( 2 \delta_\sigma^0 
c^0 \tilde f \tilde g^{0k} - \delta_\sigma^0 c^k - \delta_\sigma^k c^0 ) + \tilde g^{0\rho} \delta_\nu^k c^0 ]
\partial_\rho \bar c_\mu \partial_k ( \tilde f \delta^3),
\nonumber\\
B_{\mu\nu} &=& \tilde f ( c^0 \partial_\nu \dot{\bar c}_\mu + \tilde f \tilde g^{0\rho} \partial_\rho c^0 \partial_\nu \bar c_\mu 
+ \tilde f \tilde g^{0k} c^0 \partial_k \partial_\nu \bar c_\mu ) \delta^3 + \tilde f ( - \tilde g^{0k} c^0 + \tilde g^{00} c^k ) 
\partial_\nu \bar c_\mu \partial_k ( \tilde f \delta^3),
\nonumber\\ 
C_{\mu\nu} &=& - \tilde f^2 c^0 ( \partial_\nu \tilde g^{0\rho} \partial_\rho \bar c_\mu + \tilde g^{00} \partial_\nu \dot{\bar c}_\mu
+ \tilde g^{0k} \partial_k \partial_\nu \bar c_\mu ) \delta^3 + \tilde f ( \tilde g^{\rho k} \delta_\nu^0 - \tilde g^{0\rho} \delta_\nu^k )
\partial_k ( \tilde f c^0 \delta^3).
\label{ABC-result}  
\end{eqnarray}
Then, we see that $A_{\mu\nu} + B_{\mu\nu} + C_{\mu\nu} = 0$, thereby proving Eq. (\ref{b-b-rel}).


\end{document}